\documentclass[preprint,12pt]{elsarticle}
\usepackage{graphics}
\usepackage{graphicx}
\usepackage{subfigure}

\usepackage{latexsym}
\usepackage{amsmath}
\usepackage{graphicx}
\usepackage{times}
\usepackage{graphicx,multicol,enumerate,subfigure,amsmath}
\usepackage{apropos}

\usepackage{fullpage}
\usepackage{setspace}
\usepackage[english]{babel}
\usepackage[version=3]{mhchem}
\usepackage{amsmath, amssymb}
\usepackage{verbatim, latexsym}
\usepackage{colortbl}

\usepackage{soul}
\setstcolor{red}

\biboptions{sort&compress}
\journal{Journal of Computational Physics}

\onehalfspace

\begin{document}

\begin{frontmatter}

\title{ Mode Decomposition Methods for Flows in High-Contrast Porous Media. Part I. Global Approach}
\author{\textbf{Mehdi Ghommem}$^1$ \corref{cor1}}
\cortext[cor1]{Email address : mehdi.ghommem@kaust.edu.sa}

\author{\textbf{Victor M. Calo}$^{1}$, and \textbf{Yalchin Efendiev}$^{1,2}$ }

\address{$^{1}$ Applied Mathematics and Computational Science and Earth Sciences and Engineering \\
Center for Numerical Porous Media (NumPor) \\
King Abdullah University of Science and Technology (KAUST) \\
Thuwal 23955-6900, Kingdom of Saudi Arabia}

\address{$^{2}$ Department of Mathematics \& Institute for Scientific Computation (ISC) \\
Texas A\&M University \\
College Station, Texas, USA}

\begin{abstract}
We apply dynamic mode decomposition (DMD) and proper orthogonal decomposition (POD)
methods to flows in highly
heterogeneous porous media to extract the dominant coherent structures and derive reduced-order models via Galerkin projection.
Permeability fields with high contrast
 are considered to investigate the capability of these techniques to capture the
main flow features and forecast the flow evolution within a certain accuracy. A DMD-based approach shows
a better predictive capability due to its ability to
accurately extract
the information relevant to  long-time dynamics, in particular,
the slowly-decaying eigenmodes corresponding to largest eigenvalues.
Our study enables a better understanding of the strengths and weaknesses of the applicability of these techniques for flows in
high-contrast porous media.
Furthermore, we discuss the robustness of DMD- and POD-based reduced-order models with respect to variations in initial conditions, permeability fields,
and forcing terms.

\end{abstract}

\begin{keyword}
Model reduction, highly heterogeneous porous media, dynamic mode decomposition, proper orthogonal decomposition.
\end{keyword}

\end{frontmatter}

\section{Introduction}

In many relevant porous media engineering applications, media permeability can vary several
orders of magnitude. For example, in flow through fractured
porous media, the conductivity within fractures can be several orders
of magnitude higher than the conductivity within the matrix.
Similarly, shale barriers with very low permeability
can affect flow behavior significantly. Because of these large variations in coefficient values and the fact that these features (fractures and shale barriers)
have small spatial dimensions, the direct numerical simulations of these processes may be prohibitively expensive.
In particular, the resulting large number of degrees of freedom and associated computational cost
could inhibit the capability to perform a sensitivity analysis or conduct uncertainty quantification studies
which require many functional evaluations. As such, constructing model
reduction techniques that can judiciously select the dominant modes
corresponding to dominant flow features is important in these applications. This enables the derivation of reduced-order models with significantly less degrees of freedom while neglecting irrelevant features of the physics in order to remain computationally tractable. In this paper, we discuss global model reduction techniques for flows
in highly heterogeneous media with high contrast.

Modeling of flow in a high-contrast subsurface requires capturing its long-term dynamics
that is due to heterogeneous diffusion in the low conductivity regions.
In this paper,
our goal is to develop model reduction techniques that are suitable
for accurately predicting the behavior of flows in high-contrast media in long time scales. This is motivated by many important
applications (see \cite{EGG_MultiscaleMOR} and references therein)
where the flow response due to low conductivity
regions needs to be detected in order
to identify the location of these regions and their intensity.
The sizes of the problems involving these features are very large because shale layers or low conductivity features
can be thin and long and because of the high
contrast one needs  many grid blocks to resolve these functions. In addition, this small-scale fractures may have significant spatial variations of the coefficients. Furthermore, these problems usually need to be solved
for several initial conditions, system's settings, and forcing inputs. To account for all of these, one needs to consider robust reduced-order models that enable reliable simplified simulations.

Several techniques, such as balanced truncation, proper orthogonal decompositions (POD), and dynamic mode decomposition (DMD), have been efficiently used for global model reduction, most of which involve projection of the original governing equations onto a set of
modes. Proper orthogonal decomposition (POD) \cite{Lumley1967,Sirovich1987A,Deane1991,Berkooz1993, Holmes1996,Akhtar2009B,Wang2011A,Wang2012A,Akhtar2012A,Hay2010A,Hay2011} constitutes a common technique for extracting the coherent structures from a linear or nonlinear dynamical process. This method is based on processing information from a sequence of snapshots and identifying a low-dimensional set of basis functions that represent the most energetic structures. These functions are then used to derive a low-dimensional dynamical system that is typically obtained by Galerkin projection \cite{Akhtar2009B,Wang2011A,Wang2012A,Akhtar2012A,GhommemPCFD2012}. Schmid \cite{Schmid2010} has recently introduced a model-free decomposition technique, namely dynamic mode decomposition (DMD), to accurately extract coherent and dynamically relevant structures. This method enables the computation, from empirical data, of the eigenvalues and eigenvectors of a linear model that best represents the underlying dynamics, even if those dynamics are produced by a nonlinear process. This technique has been successfully applied for the analysis of experimental \cite{Duke2012,Schmid2011B,Pan2011,Schmid2009,Schmid2009B,Lusseyran2011} and numerical \cite{Muld2012,Schmid2011,Schmid2010,Seena2011,Mizuno2011} flow field data and has shown a great capability to capture the relevant associated dynamics.

For flows in high-contrast media, long-term dynamics are controlled by the flow in low conductivity regions. As such, it is essential to keep the modes that capture the local slow dynamical behavior in the low conductivity regions when selecting basis functions to represent the solution space and derive a reduced-order model by Galerkin projection. As for POD, the selection of modes is based on an energy ranking of the coherent structures. However, the energy may not in all circumstances be the appropriate measure to rank the importance of the flow structures and especially to detect the slow dynamics. Thus, reduced-order models generated by projection onto principal components, such as POD modes, unless selected carefully, may be inaccurate because the dominant modes identified from a set of snapshots may not necessarily correspond to the dynamically-important ones. Moreover, principal components, such as POD, -based reduced-order models are often limited in their applications due to their lack of robustness with varying flow parameters, initial conditions, and forcing inputs.

%

In this work, we apply POD and DMD approaches to flow in highly heterogeneous porous media with high contrast to identify the important physical features and derive reduced-order models that accurately capture the long-term dynamics of the system. Different numerical examples of flows in porous media characterized by varying permeability fields are considered. These permeability fields include channels and inclusions of high and low conductivity. These configurations lead to different types of behavior. The objective is to investigate the capability of POD and DMD to capture the main flow characteristics and forecast the flow dynamics response within a certain accuracy while reducing the computational cost. In all cases, the DMD-based approach shows a better predictive capability and reproduces the flow field with a more reliable accuracy. We observe convergence to small errors as the steady-state solution develops when using DMD modes while larger errors are obtained when POD modes are considered in the construction of reduced-order models. This is mostly due to the DMD's ability to extract the  information relevant to the slow dynamics in the system which control the long-time behavior. We also consider parameter-dependent problems to investigate the robustness of the POD and DMD modes with respect to variations in the initial conditions, permeability field, and forcing inputs. This analysis is motivated by many applications where one needs to solve many forward problems corresponding different permeability fields; for instance, when stochastic descriptions are used, such as when the permeability field may be subject to uncertainty or multi-phase flow where the permeability is modulated by large-scale
mobility. We show that using basis functions generated from a DMD-based approach which are determined from few selected samples, one can make accurate predictions of the dynamical behavior of the flow in highly heterogeneous porous media.

\section{Problem Formulation and Numerical Examples}
We consider a time-dependent single-phase flow in porous media governed by the following parabolic partial differential equation
\begin{eqnarray}
\frac{\partial u}{\partial t} = \nabla (\kappa(x) \nabla u) + f(x)   \quad \mbox{on} \quad \Omega \label{Eq17}
\end{eqnarray}
where $u$ is the pressure, $\Omega$ is a bounded domain, $f$ is a forcing term, $\kappa(x)$ is a positive definite scalar function that is a function of the spatial location $x$. $\kappa$ represents the ratio of the permeability over the fluid viscosity which is a highly-heterogeneous field with a high contrast (i.e., large variations in the permeability).
The high and low permeability
regions are typically heterogeneous and flow within them can be have a
complicated structure.

\begin{figure}[ht]
 \begin{center}
 \includegraphics[width=0.65\textwidth]{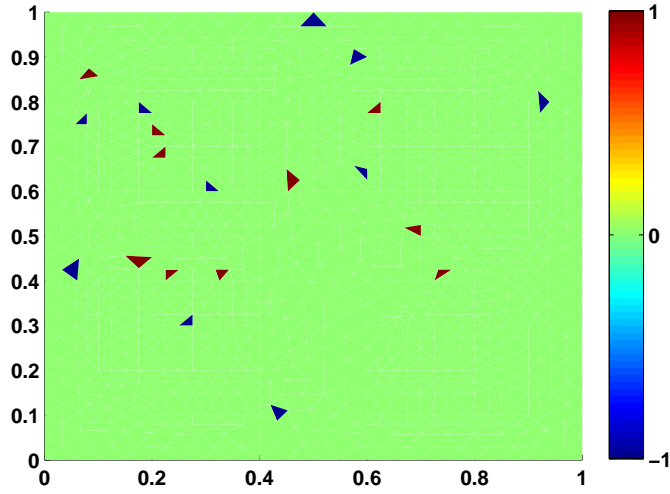}
 \end{center}
\caption{Spatial variations of the forcing $f$ over the domain $\Omega$.}\label{forcing}
\end{figure}

We consider $\Omega=]0 \; 1[\times ]0 \; 1[$, $u=0 \quad \mbox{on} \quad \partial \Omega$, and $f$ is a heterogeneous spatial field
representing injection and production rates.
 A finite element mesh is constructed by decomposing the domain $\Omega$ into $N_{elt}$ triangular elements. The variations of the forcing term $f$ in our numerical examples over $\Omega$ is shown in Figure \ref{forcing}. For each cell, the value of $f$ varies randomly between the discrete values of -1, 0, and 1. We study the flow behavior under different permeability fields that include high and low conductivity regions. We assume homogeneous Dirichlet boundary conditions in our numerical examples. This model problem is a representation of flow in a reservoir with injection modeled by the forcing term.


The finite element discretization of Equation (\ref{Eq17}) yields a system of ordinary differential equations given by
\begin{eqnarray}
\mbox{M} \dot{\textbf{U}} + \mbox{A} \textbf{U} = \textbf{F} \label{ODE}
\end{eqnarray}
where $\textbf{U}$ and $\textbf{F}$ are vectors collecting the solution values and forcing at the nodes.
Here, $\mbox{A}=(a_{ij})$, $a_{ij}=\int_\Omega \kappa \nabla \phi_i \cdot \nabla \phi_j $,
 $\mbox{M}=(m_{ij})$, $m_{ij}=\int_\Omega  \phi_i \phi_j $,
where $\phi_i$ is piecewise linear basis functions defined on
a triangulation of $\Omega$ and $\nabla$ denotes spatial gradient.

 Employing the backward Euler implicit scheme for the time marching process, we obtain
\begin{eqnarray}\label{AE}
\textbf{U}^{n+1}=\Big(\mbox{M}+\Delta t \mbox{A}\Big)^{-1}\mbox{M}\; \textbf{U}^{n} + \Big(\mbox{M}+\Delta t \mbox{A}\Big)^{-1} \Delta t\; \textbf{F}
\end{eqnarray}
where $\Delta t$ is the time step and the superscript $n$ refers to the temporal level of the solution.

\begin{figure}
  \begin{center}
      \subfigure[Inclusion - low conductivity]{\includegraphics[width=0.45\textwidth]{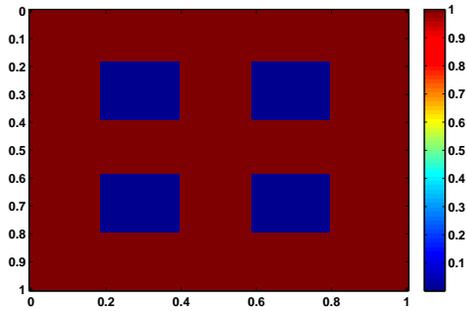}\label{Perm_incl1}}
      \subfigure[Inclusion - high conductivity]{\includegraphics[width=0.45\textwidth]{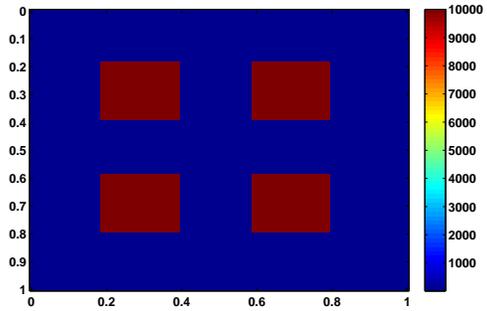}\label{Perm_incl2}} \\
      \subfigure[Channel - low conductivity]{\includegraphics[width=0.45\textwidth]{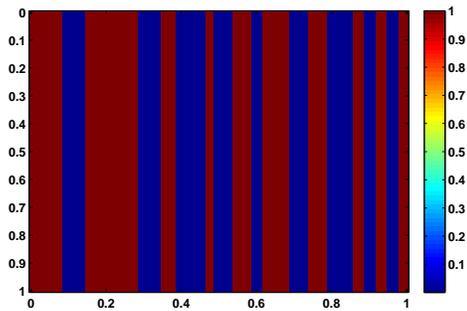}\label{Perm_cha1}}
      \subfigure[Channel - high conductivity]{\includegraphics[width=0.45\textwidth]{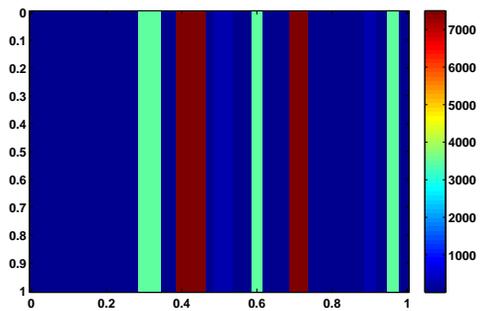}\label{Perm_cha2}}
  \end{center}
  \caption{Four different configurations of the permeability field. High and low conductivity values inside the inclusions and channels are considered while the background value is kept equal to one.}
  \label{Perm}
\end{figure}

\begin{figure}[ht]
 \begin{center}
 \includegraphics[width=0.65\textwidth]{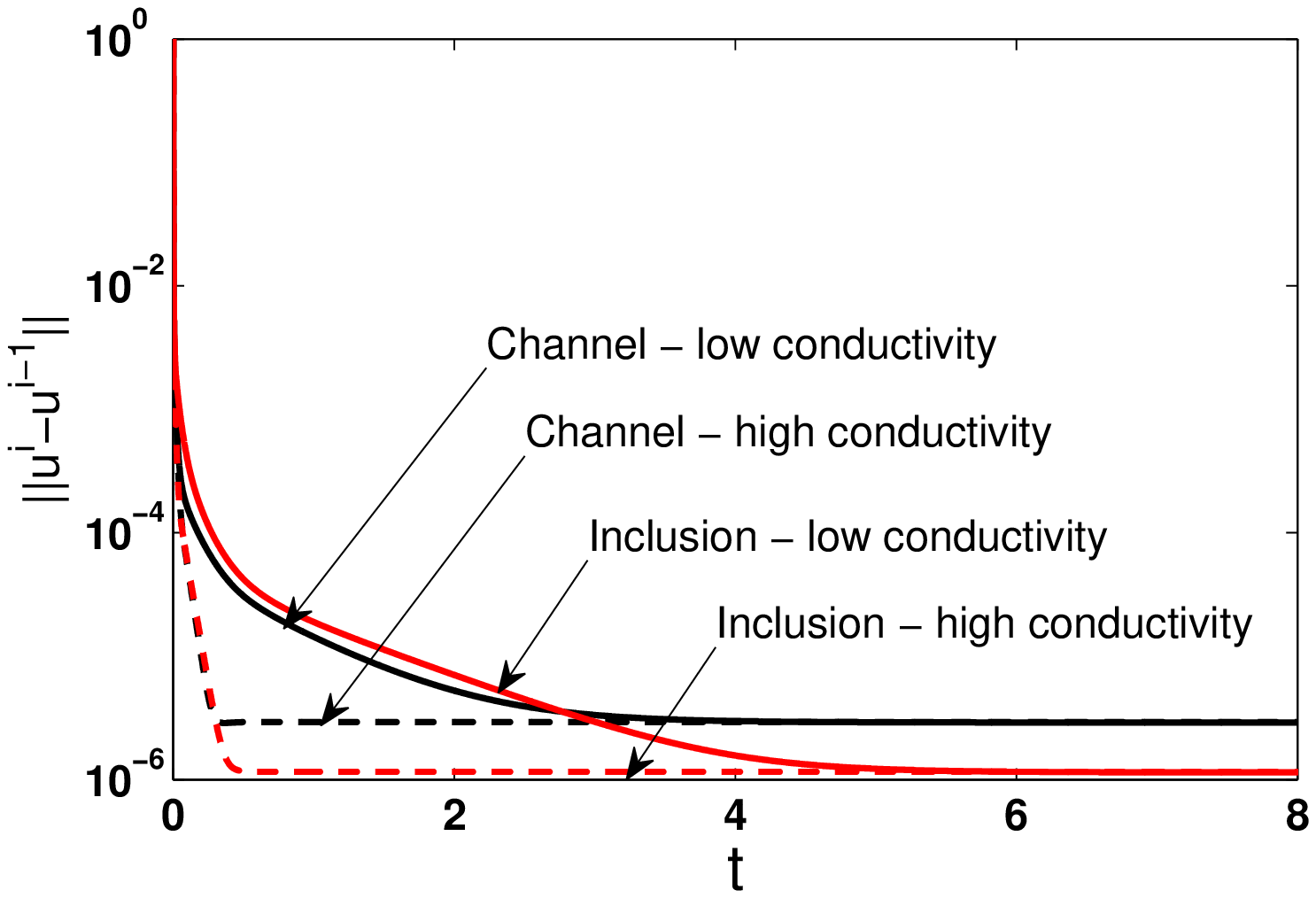}
 \end{center}
\caption{Convergence to steady-state configuration.}\label{Conv}
\end{figure}

\begin{figure}
  \begin{center}
      \subfigure[Inclusion type]{\includegraphics[width=0.45\textwidth]{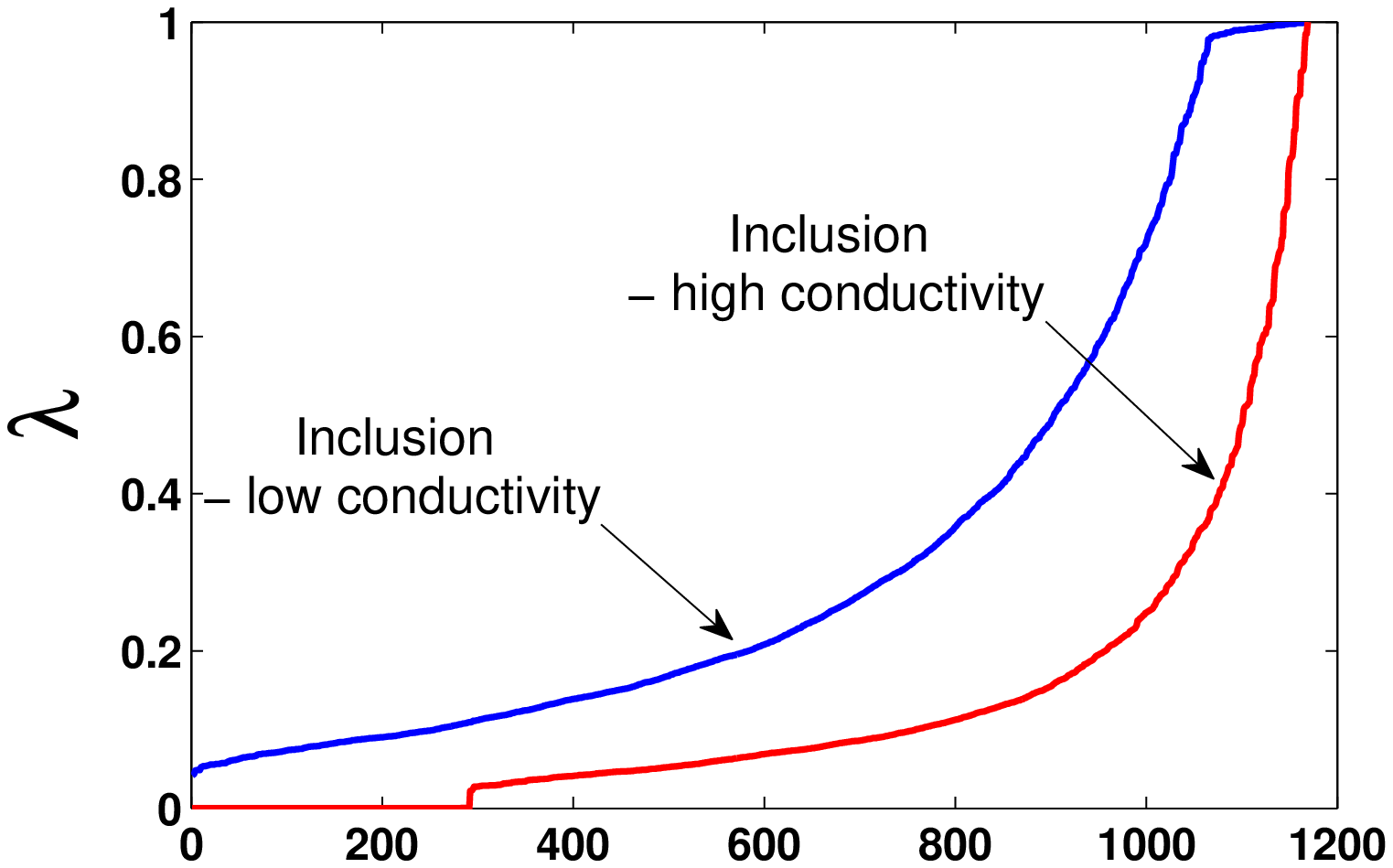}\label{Eig_inc}}
      \subfigure[Channel type]{\includegraphics[width=0.45\textwidth]{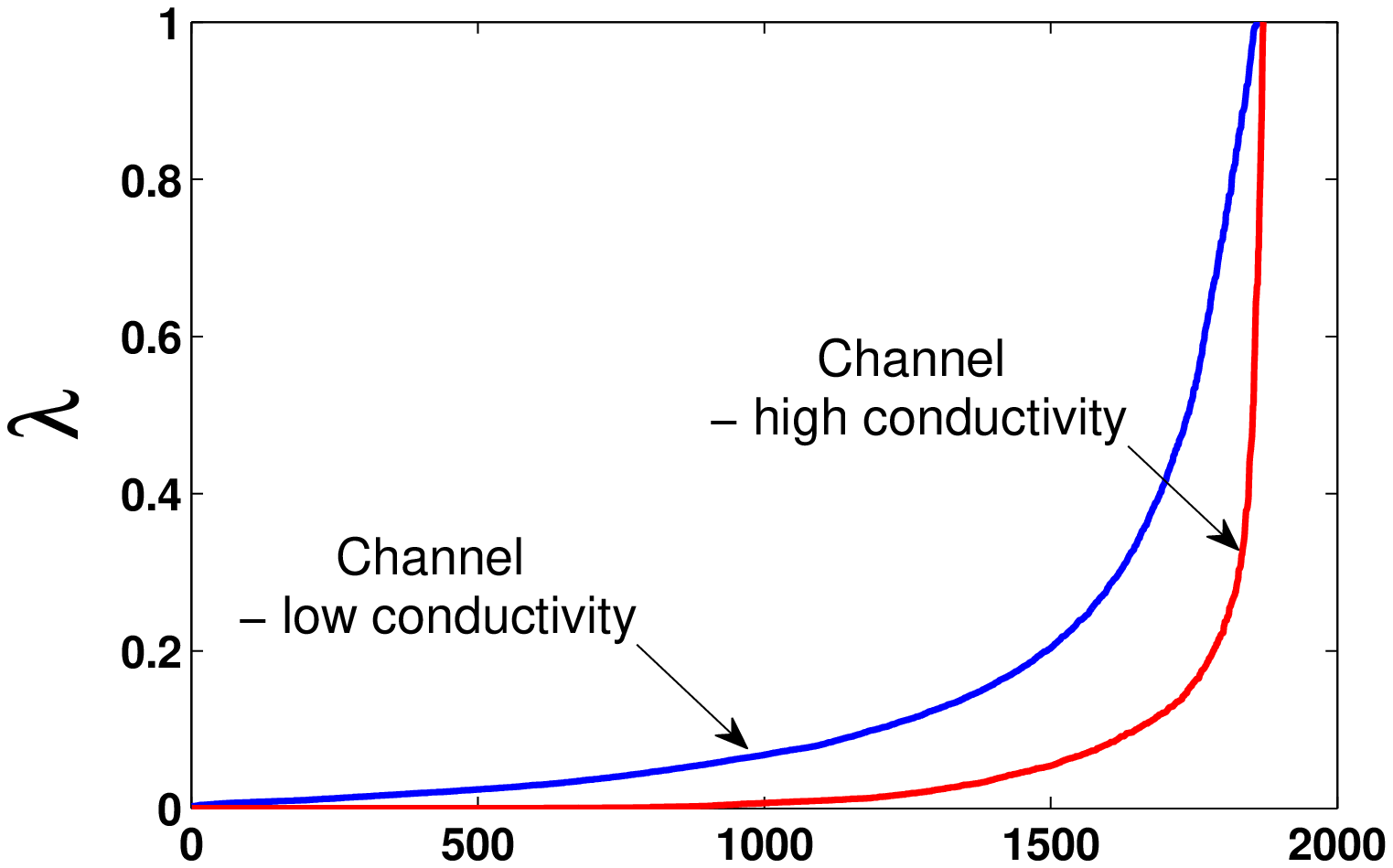}\label{Eig_incl}}
  \end{center}
  \caption{Eigenvalues of the dynamic process: (a) inclusion type permeability field and (b) channel type permeability field..}
  \label{Eig}
\end{figure}

We analyze two types of the permeability fields, namely those that contain inclusions and channels as shown in Figure \ref{Perm}. For each type, we consider two cases of high and low conductivity values inside the domain $\Omega$ while the background value  is kept equal to one. The mesh resolves the high or low conductivity regions. For all cases considered in the subsequent analysis, we assume the forcing distribution shown in Figure \ref{forcing} unless stated differently. To get an insight on the dynamical behavior of the flow under the different permeability configurations considered in this study, we plot the temporal variations of the normalized relative L$_2$ error between two successive solutions $||u^{i}-u^{i-1}||_2\;/\;||u^{i-1}||_2$ in Figure \ref{Conv}.
As expected, for the cases where  channel or inclusion permeabilities have low conductivity, the flow field takes longer time to reach the steady state.
The slow dynamics can be inferred from the eigenvalues of the matrix $\Big(\mbox{M}+\Delta t \mbox{A}\Big)^{-1}\mbox{M}$ shown in Figure \ref{Eig}. We observe that when the permeability has low conductivity regions, there are many eigenvalues that are close to $1$. The eigenmodes corresponding to these eigenvalues that are close to the unity will control long-term dynamics of the flow. The eigenvectors corresponding to these eigenvalues simply represent finite element degrees of freedom with support in the low conductivity regions. In fact, the number of these modes that are asymptotically close to one when the low conductivity parameter approaches to zero is equal to the number of degrees of freedom within low conductivity regions. This can be observed by noting that eigenvectors corresponding to small eigenvalues of Rayleigh Quotient ${\int_\Omega \kappa |\nabla \phi|^2\over \int_\Omega |\phi|^2}$ (that corresponds to eigenvalues which are close to the one) consist of functions that vary within low conductivity regions while constant in high conductivity regions.

\subsection{Mode decomposition methods}
In the subsequent analysis, we present numerical results to demonstrate the capability of the proper orthogonal decomposition (POD) and the dynamic mode decomposition (DMD) techniques to reconstruct the flow field. First, we provide a brief description of POD and DMD and their implementation. We then discuss the usefulness and effectiveness of these decomposition methods to capture the relevant flow characteristics and derive reduced-order models that will be used to predict the flow field under different configurations obtained by varying the initial conditions and permeability field.

\subsubsection{Proper Orthogonal Decomposition}\label{POD}
A classical way to compute POD modes is to perform a singular value decomposition (SVD) of the algebraic operator that maps the states between different realizations, however, this approach may have a limited application, especially when dealing with a large mesh size as in direct numerical simulations (DNS) for which fine meshes (high resolution) imply high computational
cost. Alternatively, one could use the method of snapshots \cite{Sirovich1987B} which allows for a significant reduction of the
large data sets. In this method, sets of instantaneous solutions (or snapshots) of the flow parameters obtained from the DNS are generated and stored
in an $M-\mbox{by}-N$ matrix $C$ where $M$ and $N$ denote, respectively, the number of grid points and snapshots.  Since $M\gg N$, we seek to compute the singular values of $C$ as well as its right singular vector matrix through an eigen analysis of the matrix $C^* C$; that is,

\begin{eqnarray}
C^*\; C\in \mathbf{R}^{N\times N}:C^*\; C \; V_i=\sigma_i^2V_i\quad \mbox{and}\quad
\phi^{POD}_i=\frac{1}{\sigma_i}C\; V_i
\label{eqn:CRmat}
\end{eqnarray}
where $\sigma_i$ are referred to singular values and $V_i$ are the eigenvectors of the matrix $C^*\; C$.

The selection of POD modes is optimal in the sense that the error between each snapshot and its projection on the space spanned by those modes is minimized \cite{Burkardt}. Besides, the square of the singular values represents a measure of the energy content of each POD mode and thus provides
guidance for the number of modes that should be considered in order to capture the relevant physics of the system.

\subsubsection{Dynamic Mode Decomposition }

The basic principles and mathematical background of dynamic mode decomposition (DMD) are given below following \cite{Schmid2010,Schmid2011}. Over the last few years, this technique has been widely applied on experimental \cite{Duke2012,Schmid2011B,Pan2011,Schmid2009,Schmid2009B,Lusseyran2011} and numerical \cite{Muld2012,Schmid2011,Schmid2010,Seena2011,Mizuno2011} flow field data to identify dominant coherent structures and help in understanding the underlying physics. These structures can be used to project a large-scale problem onto a low-dimensional subspace to obtain a dynamical system with much fewer degrees of freedom. For a more detailed description, the reader is referred to \cite{Schmid2010,Schmid2011,Schmid2011B,Chen2012}.

The DMD method is based on postprocessing a sequence of snapshots to extract the dynamic information. Let a snapshot sequence, separated by a constant time step $\Delta t$, collected in a matrix $\textbf{V}_1^N$; that is,
\begin{eqnarray}
\textbf{V}_1^N=\{\textbf{v}_1,\textbf{v}_2,\textbf{v}_3,\cdots, \textbf{v}_N\}
\label{Eq1}
\end{eqnarray}
where $\textbf{v}_i$ denotes the $i^{\mbox{th}}$ solution field and the subscript denotes the first element of the sequence while the superscript denotes the last element. In Equation (\ref{Eq1}), these correspond to $1$ and $N$, respectively. The basic idea of DMD is to relate the solution field $\textbf{v}_i$ to the subsequent solution field $\textbf{v}_{i+1}$ through a linear mapping $\textbf{A}$; that is,
\begin{eqnarray}
\textbf{v}_{i+1}=\textbf{A}\textbf{v}_i. \label{Eq2}
\end{eqnarray}
This assumption leads to a representation of the solution field as a Krylov sequence
 \begin{eqnarray}
\textbf{V}_1^N=\{\textbf{v}_1,\textbf{A}\textbf{v}_1,\textbf{A}^2\textbf{v}_i,\cdots, \textbf{A}^{N-1}\textbf{v}_1\} \label{Eq3}
\end{eqnarray}
The objective is to determine the main characteristics of the dynamical process represented by the linear mapping $\textbf{A}$ (even if the solution field involves nonlinear aspects). This is performed by computing (or approximating) the eigenvectors and eigenvalues of the matrix $\textbf{A}$. For a very large system, these computations may be numerically intractable. Furthermore, for instance in experimental fluid simulations, the exact form of the matrix $\textbf{A}$ is not given. As such, an efficient and fast numerical approach that approximates well the slow dynamics is useful.

Assuming that for sufficiently long sequence, the vector $\textbf{v}_N$ can be represented by a linear combination of the previous solution fields; that is,
\begin{eqnarray}
\textbf{v}_N = \sum_{i=1}^{N-1} a_i \textbf{v}_i + \textbf{r} \label{Eq4}
\end{eqnarray}
or
\begin{eqnarray}
\textbf{v}_N = \textbf{V}_1^{N-1} \; \textbf{a} + \textbf{r} \label{Eq5}
\end{eqnarray}
where $\textbf{a}=\{a_1,a_2,\cdots,a_{N-1}\}^*$ and $\textbf{r}$ is the residual vector.
Combining Equations (\ref{Eq3}) and (\ref{Eq5}) and rearranging the result, we obtain
\begin{eqnarray}
\textbf{A}\; \textbf{V}_1^{N-1} = \textbf{V}_2^{N} = \textbf{V}_1^{N-1} \; \textbf{S}+ \textbf{r} \; \textbf{e}^*_{N-1} \label{Eq6}
\end{eqnarray}
where $\textbf{e}^*_{N-1}=\left(
                            \begin{array}{cccc}
                              0 & \cdots & 0 & 1 \\
                            \end{array}
                          \right)
$ is the $(N-1)$ unit vector and the matrix $\textbf{S}$ is a companion matrix defined as:
\begin{eqnarray}
\textbf{S} =\left(
                   \begin{array}{ccccc}
                     0 &  &  &  & a_1 \\
                     1 & 0 &  &  & a_2 \\
                      & \ddots & \ddots &  & \vdots \\
                      &  & 1 & 0 & a_{N-2} \\
                      &  &  & 1 & a_{N-1} \\
                   \end{array}
                 \right) \label{Eq7}
\end{eqnarray}

The unknown matrix $\textbf{S}$ is determined by minimizing the residual $\textbf{r}$ which is obtained by expressing the $N^{th}$ snapshot $\textbf{v}_N$ by a linear combination of $\{\textbf{v}_1,\textbf{v}_2,\textbf{v}_3,\cdots, \textbf{v}_{N-1}\}$ in a least-squares sense. The minimization problem to determine $\textbf{S}$ is expressed then as:
\begin{eqnarray}
\textbf{S}=\min_{\textbf{S}} \parallel \textbf{V}_2^{N} - \textbf{V}_1^{N-1}\textbf{S} \parallel \label{Eq8}
\end{eqnarray}
To avoid cumbersome notation, we use the same variable for the minimizer as
the variable that is being minimized.
The solution can be determined either using a QR-decomposition of the snapshot matrix $\textbf{V}_1^{N-1}$; that is \cite{Schmid2011},
\begin{eqnarray}
\textbf{S}= \textbf{R}^{-1} \; \textbf{Q}^* \; \textbf{V}_2^{N} \label{Eq9}
\end{eqnarray}
where the matrices $\textbf{Q}$ and $\textbf{R}$ are obtained from QR decomposition of the snapshot matrix $\textbf{V}_1^{N-1}$ or using the Moore-Penrose pseudoinverse of the matrix $\textbf{V}_1^{N-1}$ to obtain a solution \cite{Chen2012}
\begin{eqnarray}
\textbf{S}= \Big((\textbf{V}_1^{N-1})^*\;\textbf{V}_1^{N-1}\Big)^{-1} \; (\textbf{V}_1^{N-1})^* \; \textbf{V}_2^{N} \label{Eq11}
\end{eqnarray}
Once $\textbf{S}$ is computed, the next step is to evaluate its eigenvalues and eigenvectors collected in the diagonal matrix $\textbf{D}$, the matrix $\textbf{X}$, respectively.

Finally, the DMD spectrum $\lambda_j$ is obtained by transforming the eigenvalues of $\textbf{S}$ from the time-stepper format to the format more commonly
used in stability theory (accomplished through a logarithmic mapping), and the dynamic modes are computed $\phi^{DMD}_j$ by weighing the snapshot based by the
eigenvectors of \textbf{S}; that is,
\begin{eqnarray}
\lambda_j = \log(\textbf{D}_{jj}) / \Delta t \label{Eq13}
\end{eqnarray}
\begin{eqnarray}
\phi^{DMD}_j = \textbf{V}_1^{N-1} \; \textbf{X}_j \label{Eq14}
\end{eqnarray}
where $\textbf{X}_j$ is the $j^{\mbox{th}}$ column of the matrix $\textbf{X}$.

The above algorithm to compute the dynamic modes based on the companion matrix $\textbf{S}$ may be ill-conditioned in practical situation. As such,
Schmid \cite{Schmid2010} proposed a more robust implementation. The algorithm includes the following steps.
First, we let $\textbf{V}_1^{N-1} = \textbf{U} \; \Sigma \; \textbf{W}^* $ be the singular value decomposition of the data sequence $\textbf{V}_1^{N-1}$. Substituting the SVD representation $\textbf{U} \; \Sigma \; \textbf{W}^* $ into Equation (\ref{Eq6}) and multiplying the result by from the left $\textbf{U}^*$ and by $\textbf{W} \; \Sigma^{-1}$ from  the right, we obtain the following matrix
\begin{eqnarray}
\textbf{U}^* \; \textbf{A} \; \textbf{U} = \textbf{U}^* \; \textbf{V}_2^{N} \; \textbf{W} \; \Sigma^{-1} \equiv \tilde{\textbf{S}} \label{Eq15}
\end{eqnarray}
For configuration sets with $\mbox{dim}(\textbf{v}_i)\gg N $, the method of snapshots, as described in Section \ref{POD}, can be used to avoid the computational burden associated
with the singular value decomposition of large matrices. The matrix $\textbf{U}$ contains the POD modes of the sequence of snapshots $\textbf{V}_1^{N-1}$, one may conclude that the matrix $\tilde{\textbf{S}}$ is obtained from the projection of the linear operator $\textbf{A}$, which is used to approximate the underlying dynamical process, onto
a the POD basis. A key advantage of the above implementation is the ability to take into account the rank-deficiency of $\textbf{V}_1^{N-1}$ by considering
a limited basis $\textbf{U}$ given only by the non-zero singular values of $\Sigma$ (or by singular values above a threshold that can be determined based the amount of cumulative energy that needs to be captured). The dynamic modes are computed from the matrix $\tilde{\textbf{S}}$ as follows:
\begin{eqnarray}
\phi^{DMD}_j = \textbf{U} \; \textbf{y}_j, \label{Eq16}
\end{eqnarray}
where $\textbf{y}_j$ is the $j^{\mbox{th}}$ eigenvector of $\tilde{\textbf{S}}$, i.e., $\tilde{\textbf{S}}\; \textbf{y}_j =\mu_j \textbf{y}_j$, and $\textbf{U}$ is the matrix collecting the right singular vectors of the snapshot sequence $\textbf{V}_1^{N-1}$.

In a recent paper, Chen et al. \cite{Chen2012} proposed an optimized version of the DMD algorithm where they employ a global optimization technique to minimize the residual error at all snapshots instead of the error at only the last snapshot. They tested the approach over a variety of fluid problems and showed its superiority in capturing the relevant frequencies and reproducing flowfields with small projection errors.

\subsection{Numerical Examples}
\paragraph{Pre-processing} We solve numerically Equation (\ref{Eq17}) over a time interval of $[0 \; 8000  \Delta t]$ where $\Delta t = 0.001$. This time interval is observed to be long enough to reach the steady-state solution for all cases considered, as shown in Figure \ref{Conv}. Then, we record the first 25 instantaneous solutions (usually referred as snapshots) and collect them in a snapshot matrix as:
 \begin{eqnarray}
\textbf{V}_1^N=\{\textbf{v}_1,\textbf{v}_2,\textbf{v}_3,\cdots, \textbf{v}_N\} \label{Snapm}
\end{eqnarray}
where $N$ is the number of snapshots and the size of the column vectors $\textbf{v}_i$ is denoted by $M$. In these numerical simulations, $N$ is taken equal to 25 and $M$ is equal to 1169 and 1871 for the inclusion- and channel-type permeability fields, respectively. The snapshot selection process is determinant for the accuracy of the resulting reduced-order model. The use of 25 snapshots is observed to be appropriate to reproduce results with acceptable accuracy as will be shown later.

The POD and DMD basis vectors are computed by applying the numerical approaches as described in the previous section to data obtained from flow simulations in highly heterogeneous porous media. For the sake of comparison, we extract and keep the same number of modes for POD and DMD in our simulations.

\begin{figure}
  \begin{center}
      \includegraphics[width=0.65\textwidth]{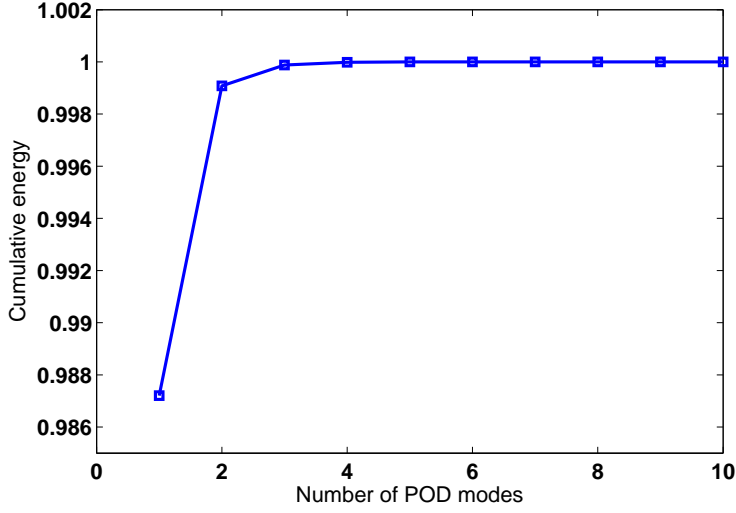}
  \end{center}
  \caption{Variations of the normalized cumulative energy content with the number of POD modes.}
  \label{Cumulative_energy}
\end{figure}

The POD technique identifies the most energetic structures. The POD is based on an energy ranking of the coherent structures obtained by enforcing the orthogonality of the correlation spatial matrix. This energy ranking is given by the singular values of the spatial correlation matrix shown in Equation (\ref{eqn:CRmat}). Thus, to gain insight into the contribution of each mode to the total energy of the system, we define the cumulative energy as $E=\sum_{i=1}^m \sigma_i$ and the cumulative contribution of the first $j$ modes as $c_j=\Big(\sum_{i=1}^j \sigma_i\Big)/E$. We plot in Figure \ref{Cumulative_energy} the variations of the normalized cumulative energy content with the number of POD modes. Most of the energy is contained in the first few modes. Specifically, the first six modes contain more than 99.9\% of the total flow energy. As such, the following numerical results from the POD and DMD-based approaches use the first six modes.

\paragraph{Approximate solution} We postprocess the snapshot matrix, as described in section 2.1, to compute the POD and DMD modes and use these modes to approximate the solution field. As such, we assume an expansion in terms of the modes $\phi^k_i$; that is, we let
\begin{equation}
u(x,t) \approx \tilde{u}(x,t) = \sum_{i=1}^{m} \alpha_i(t) \phi^k_i(x)\label{utilda}
\end{equation}
or in a matrix form
\begin{equation}
\textbf{U}^{n} \approx \tilde{\textbf{U}}^n = \Phi \alpha^n \label{utildam}
\end{equation}
where $\Phi=\left(
              \begin{array}{ccc}
                \phi_1^k & \cdots & \phi_m^k \\
              \end{array}
            \right)
$ and $k$ can refer to either POD or DMD.
\paragraph{L$_2$ projection error} To assess the capability of the POD and DMD modes in capturing the dynamics involved in the process and enabling good projection subspaces, we project each snapshot onto the POD modes, and compute the following inner product
\begin{equation}
\alpha_i(t_j)=\left< \phi^k_i(x),u(x,t_j)\right>, \;\;\mbox{or}\;\; \alpha^j=(\Phi^*\Phi)^{-1}\Phi^*\textbf{U}^j  \label{Eq19}
\end{equation}
where $\left< F,G\right>=\int_{\Omega}(F\; G) \;d\Omega$, and define the relative error as the L$_2$-norm of the difference between the exact and approximate solutions over the exact one; i.e.
\begin{equation}
\parallel E(t) \parallel_2 = \frac{\parallel \tilde{u}(x,t) - u(x,t) \parallel_2}{\parallel u(x,t) \parallel_2}. \label{Eq20}
\end{equation}

\begin{figure}
  \begin{center}
      \subfigure[Inclusion - low conductivity]{\includegraphics[width=0.48\textwidth]{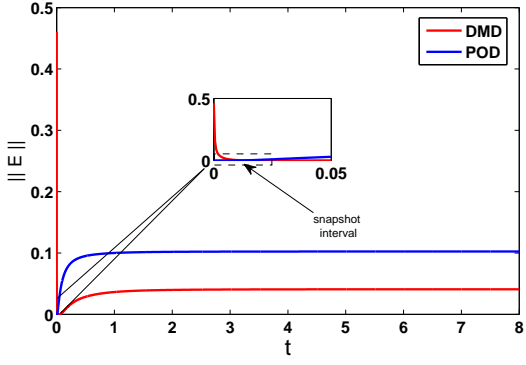}\label{ErrProj1}}
      \subfigure[Inclusion - high conductivity]{\includegraphics[width=0.48\textwidth]{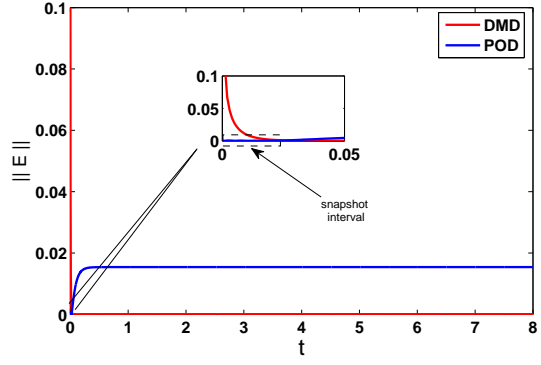}\label{ErrProj2}}\\
      \subfigure[Channel - low conductivity]{\includegraphics[width=0.48\textwidth]{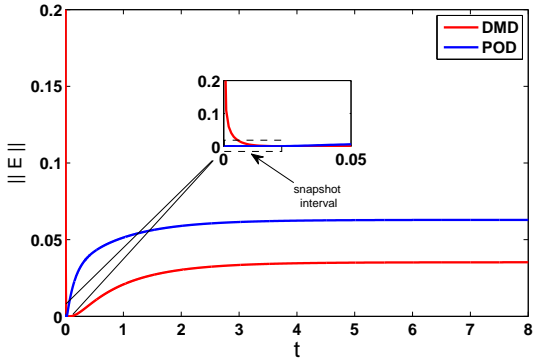}\label{ErrProj3}}
      \subfigure[Channel - high conductivity]{\includegraphics[width=0.48\textwidth]{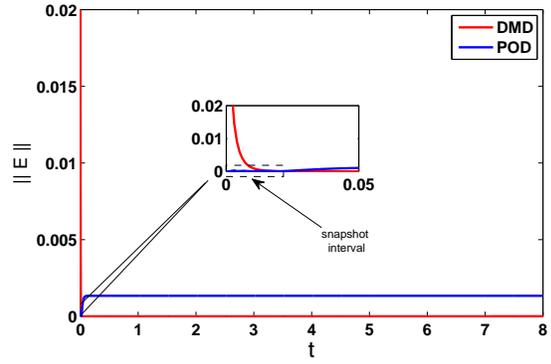}\label{ErrProj4}}
  \end{center}
  \caption{Variations of the L$_2$ projection error with time for different permeability configurations. Results are obtained using POD and DMD modes.}
  \label{ErrorProj}
\end{figure}

The L$_2$ projection error is computed for both DMD and POD- based representations. The number of modes kept in the expansion given by Equation (\ref{utilda}) is taken equal to six. The corresponding results are presented in Figure \ref{ErrorProj}. A low projection error is obtained in the interval of snapshots used to compute the modes. This is expected since POD produces the optimal modal subspace that contains the largest amount of energy. When using DMD modes to construct the projection subspace, a large error is obtained at the first few time steps and recovers as time evolves to reach small values. The L$_2$ projection error increases with time outside snapshots sequence interval and converges to steady-state values. We observe that the steady-state error values obtained when approximating the solution field in terms of the POD modes are larger than those obtained when projecting the solution field onto the space spanned by the DMD modes. These observations show that DMD performs better than POD in extracting relevant dynamic information and then has better predictive capability of the long-term dynamical behavior. However, low projection error indicates the ability of modal decomposition techniques to compute good projection subspaces, but it does not guarantee that a reduced-order model obtained mainly by Galerkin projection of the original governing equation on the subspace spanned by the modes will certainly reproduce the reference solution.

\paragraph{Reduced-order model} To obtain the reduced-order model, we use the solution expansion given by Equation (\ref{utilda}), substitute it into Equation (\ref{Eq17}) and project the result onto the space formed by the modes as
\begin{equation}\label{projection}
\left< \phi_k,\frac{\partial u}{\partial t} - \nabla \; (\kappa(x) \; \nabla u) - f\right>=0,
\end{equation}
one obtains a set of $m$ ordinary differential equations that constitute a reduced-order model; that is,
\begin{equation}\label{rom}
\dot{\alpha} = - (\Phi^*\mbox{M}\Phi)^{-1} \Phi^* \mbox{A} \Phi \alpha + (\Phi^*\mbox{M}\Phi)^{-1} \Phi^* \mbox{F}
\end{equation}

Thus, the original problem with $M$ degrees of freedom is reduced to a dynamical system with $m$ dimensions where $m \ll M$. We solve numerically Equations (\ref{rom}) and compute the approximate solution given by Equation (\ref{utilda}). Again, six modes (i.e., $m=6$) have been used. We consider the permeability field shown in Figure \ref{Perm_incl1} and run flow simulations where we vary the number of snapshots and compute the errors between the reference and approximate solutions obtained by employing POD- and DMD-based approaches. The results are presented in the Table \ref{Err_N}. Clearly, the selection of snapshots is an important step when applying POD and DMD for model reduction. Using only seven snapshots yields large value for the error while considering more snapshots improves significantly the forecasting capability of the reduced-order model and leads to small errors. The results show that POD and DMD modes computed from 25 snapshots contain relevant information of the flow dynamics and then enable a reduced-order model that predicts the flow behavior with good accuracy. Increasing the number of snapshots would decrease more the error but the aim of this study is to show the potentiality of DMD and/or POD techniques to detect the dominant modes that govern long-term dynamics from a small set of snapshots and to forecast the evolution of the flow field an acceptable accuracy.

\begin{table}[ht!]
\caption{Variations of the error at $t=5$ with the number of snapshots $N$. Results are obtained using POD and DMD modes.}\label{Err_N}
\begin{center}
\begin{tabular}{l c c c c c c}
  \hline
  \hline
  $N$ & 7 & 10 & 15 & 20 & 25 & 30 \\
  \hline
  $\parallel E \parallel_{POD}$ & 51.63 $\%$ & 36.8 $\%$& 19.59 $\%$ & 10.9 $\%$ & 7.9 $\%$ & 7 $\%$ \\
 $\parallel E \parallel_{DMD}$ & 37.6 $\%$& 7.54 $\%$& 4.69 $\%$ & 3.78 $\%$ & 2.8 $\%$ & 2.9 $\%$ \\
  \hline
  \hline
\end{tabular}
\end{center}
\end{table}

\begin{figure}
  \begin{center}
      \subfigure[Inclusion - low conductivity]{\includegraphics[width=0.48\textwidth]{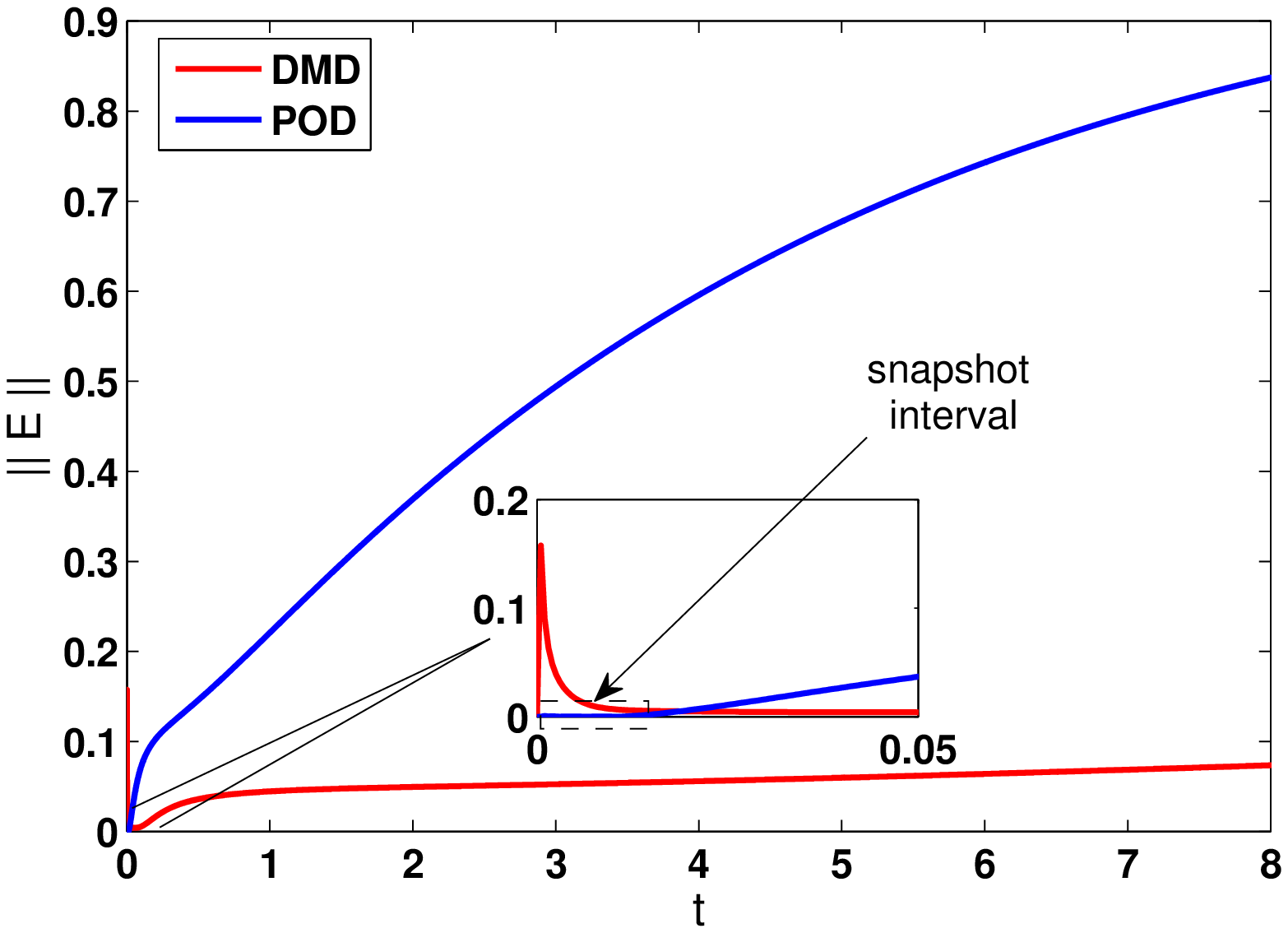}\label{ErGaler1}}
      \subfigure[Inclusion - high conductivity]{\includegraphics[width=0.48\textwidth]{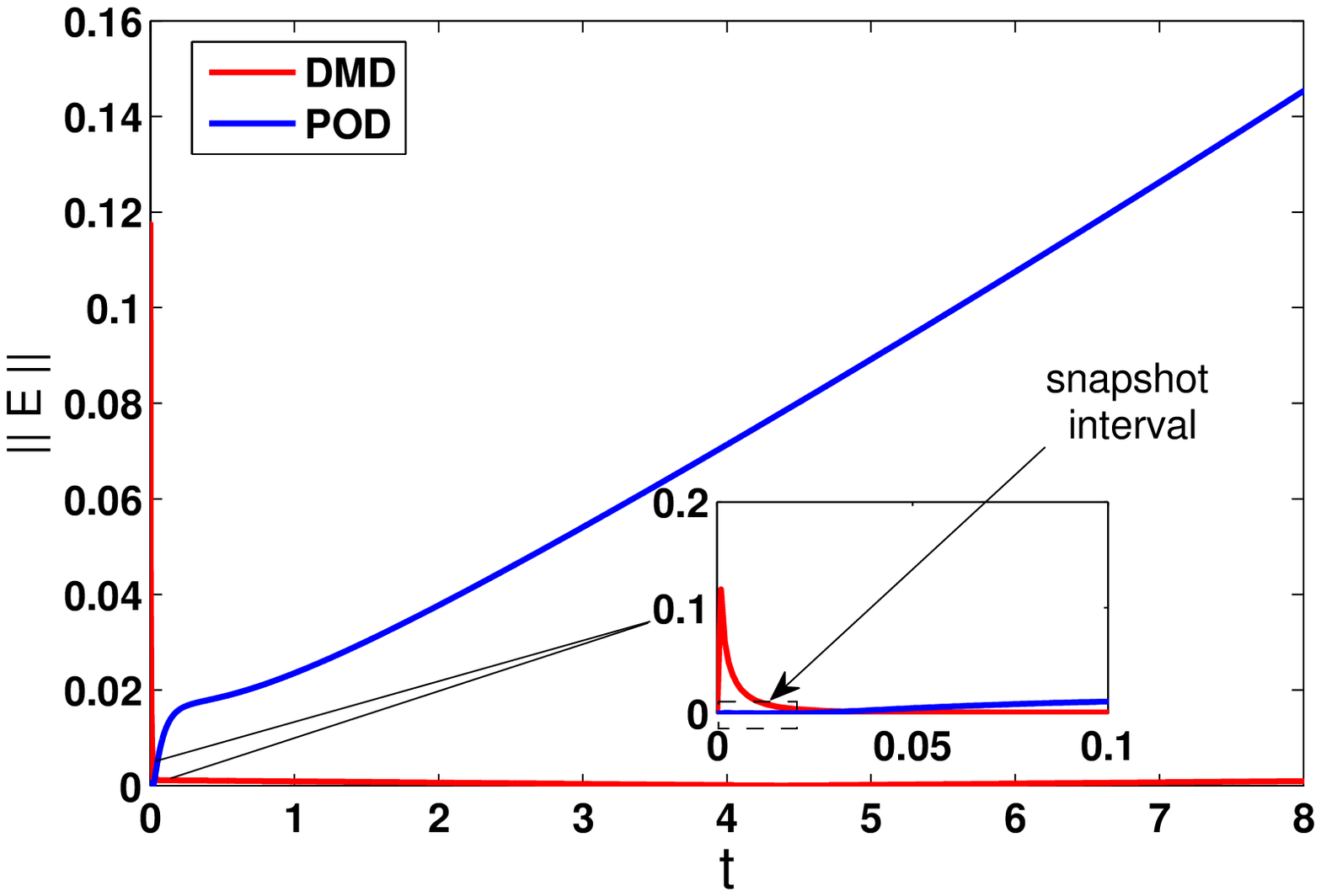}\label{ErGaler2}}\\
      \subfigure[Channel - low conductivity]{\includegraphics[width=0.48\textwidth]{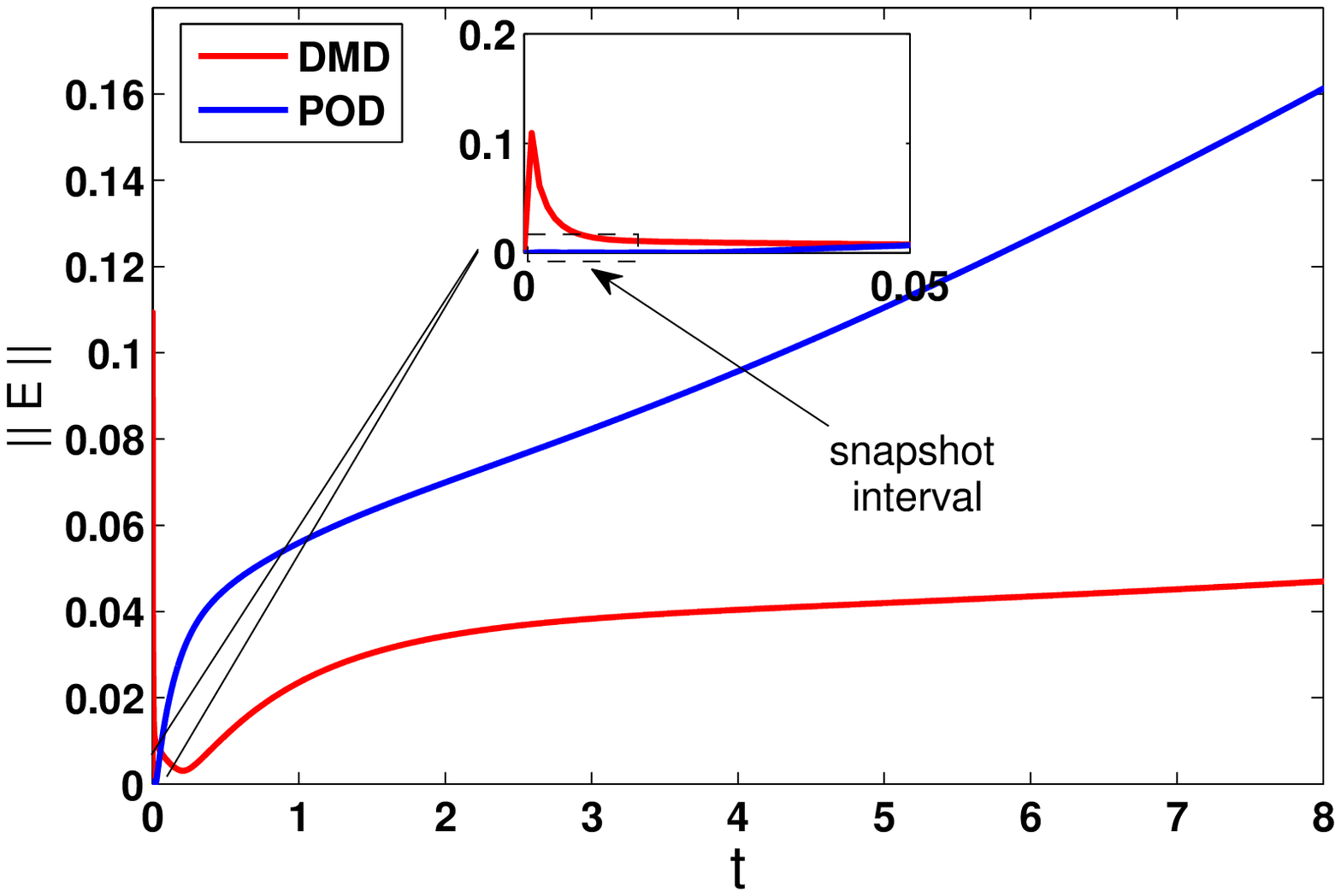}\label{ErGaler3}}
      \subfigure[Channel - high conductivity]{\includegraphics[width=0.48\textwidth]{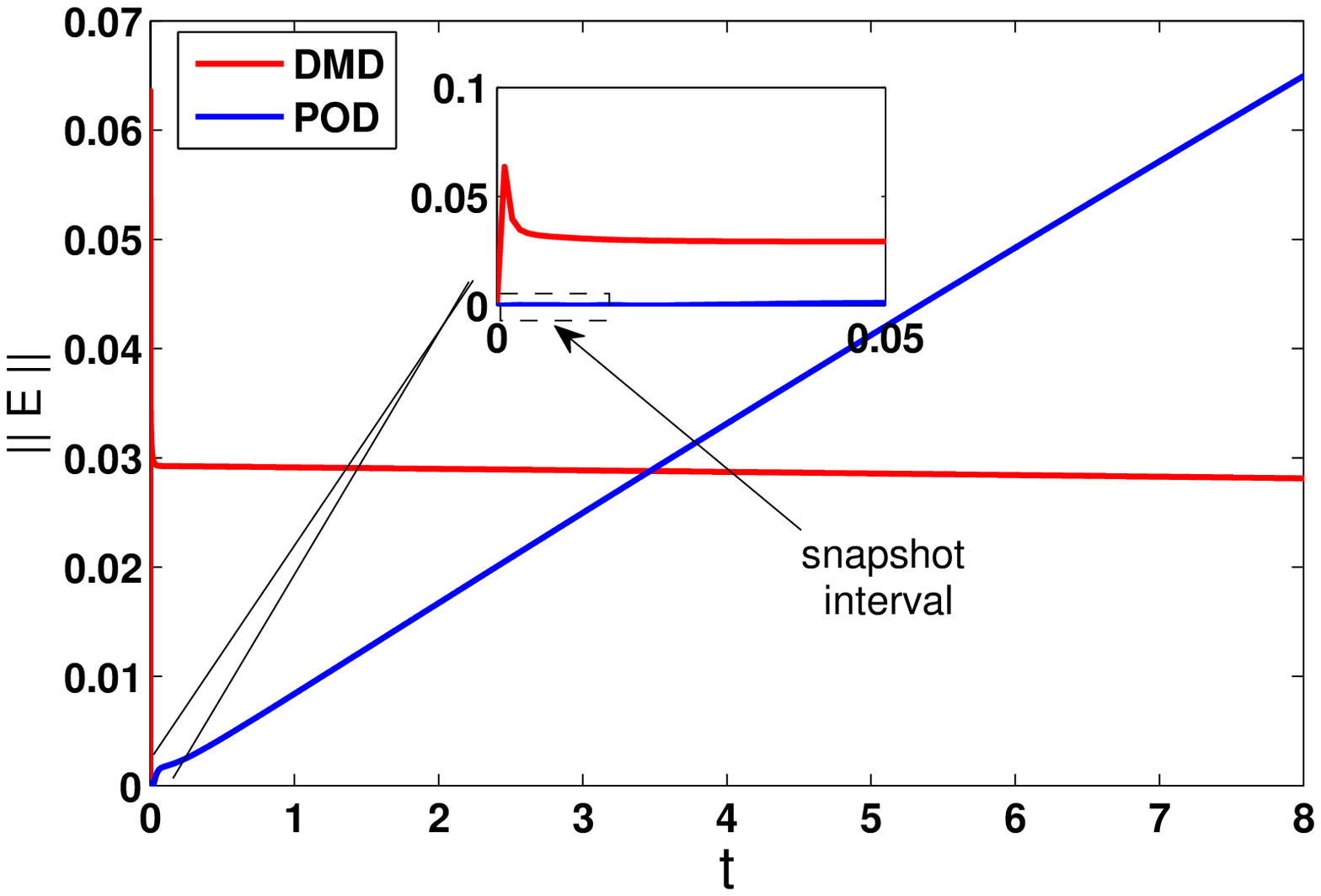}\label{ErGaler4}}
  \end{center}
  \caption{Variations of the Galerkin projection error with time for different permeability configurations. Results are obtained using POD and DMD modes.}
  \label{ErrorGaler}
\end{figure}

The variations of the projection error, as defined in Equation (\ref{Eq20}), with time for different permeability configurations are depicted in Figure \ref{ErrorGaler}. Clearly, the large projection steady-state errors (more than 80\% for the case of the inclusion permeability structure with low conductivity) obtained when deriving a reduced-order using POD modes shows that POD is able to reproduce well the flow field only within the snapshot interval while it obviously fails to predict it outside that interval. On the other hand, small projection errors between the reference and approximate solutions (except those observed at the first few time steps), as shown in Figure \ref{ErrorGaler}, demonstrates the capability of the reduce-order model obtained by projection the governing equations onto the space spanned by the DMD modes to predict the flow field within an acceptable accuracy (at most, the error reaches 8\%). Clearly, the eigenvalues and eigenvectors of the low-dimensional subspace of DMD modes capture the principal dynamics of the flow. In particular, the DMD -based approach allows for eliminating the eigenvectors associated with small asymptotically vanishing eigenvalues and detect the slow dynamics. The large errors obtained from DMD-based analysis and observed at the first few time steps are mostly due to the fact the DMD modes do not contain the eigenvectors that correspond to small eigenvalues and which govern the fast decaying dynamics.

\begin{figure}
  \begin{center}
      \subfigure{\includegraphics[width=0.33\textwidth]{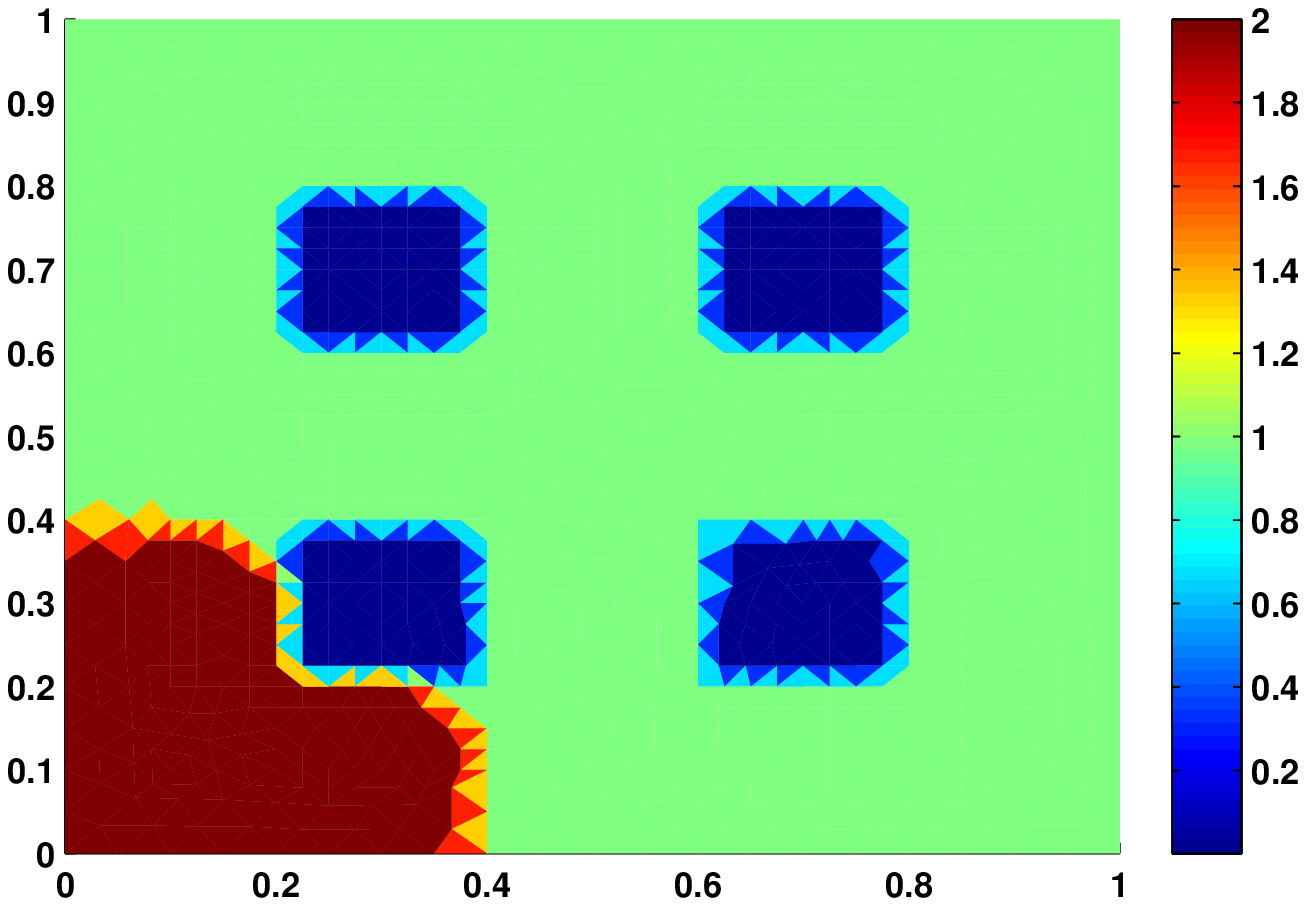}\label{tv2}}
      \subfigure{\includegraphics[width=0.32\textwidth]{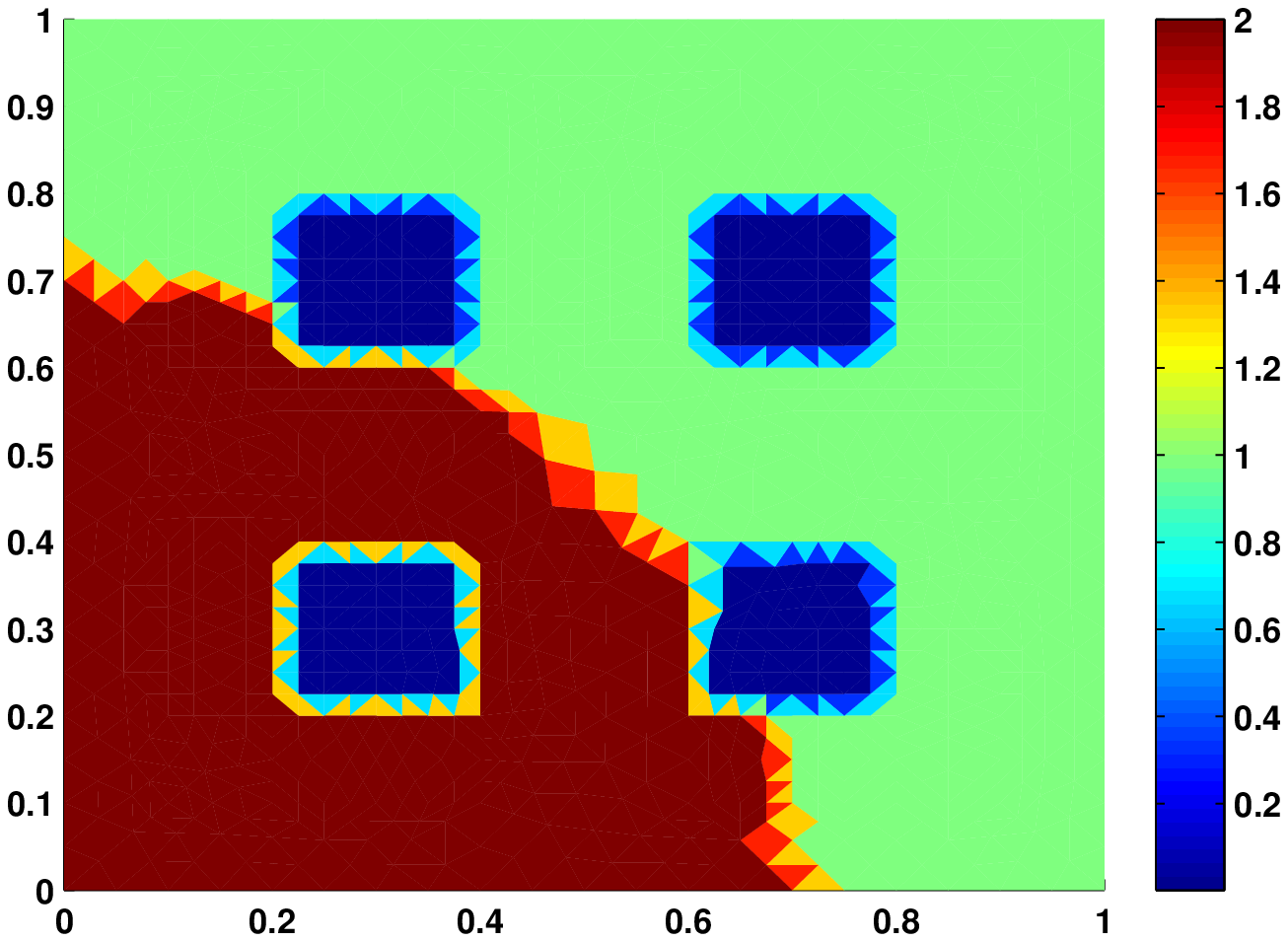}\label{tv3}}
      \subfigure{\includegraphics[width=0.32\textwidth]{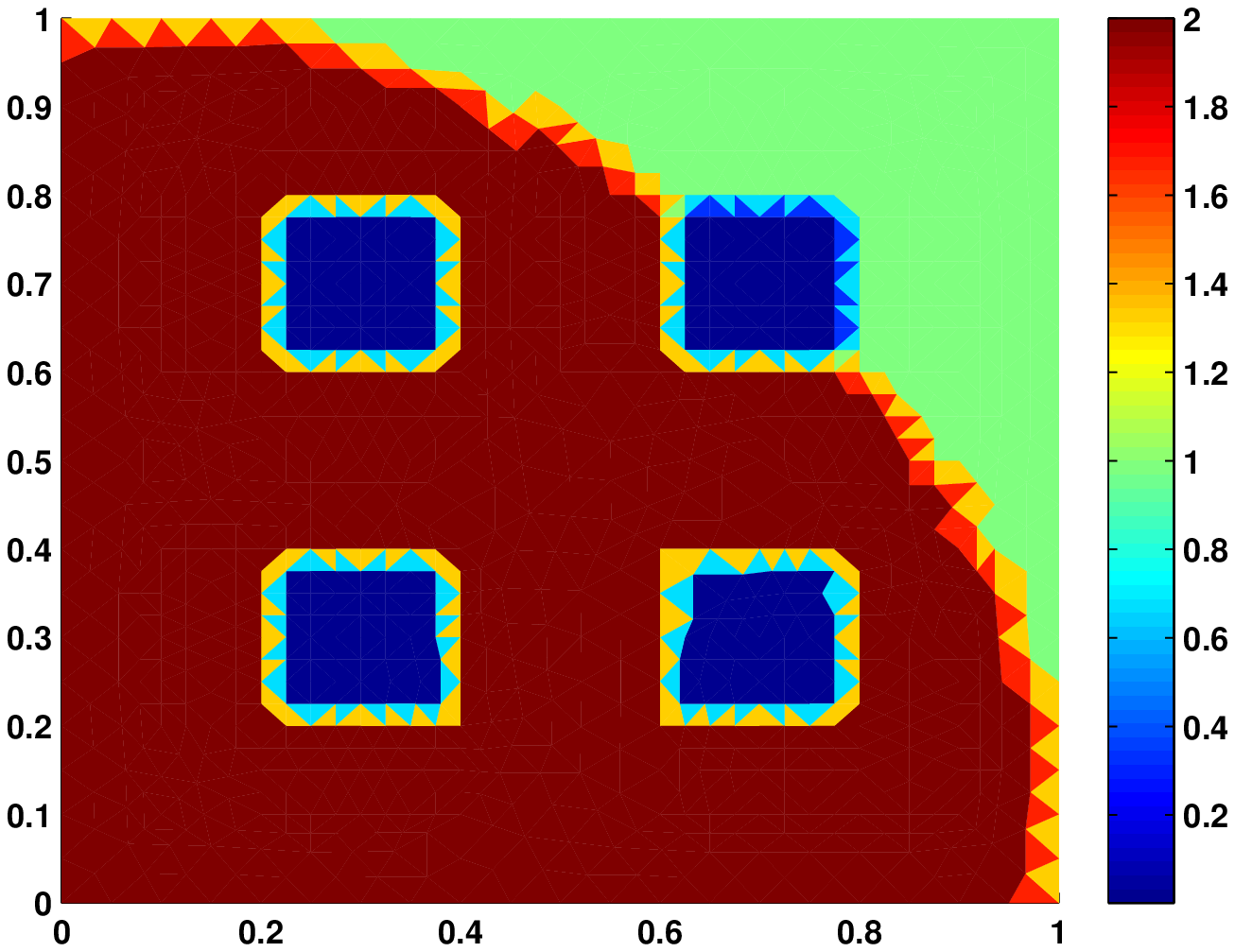}\label{tv4}}
  \end{center}
  \caption{Snapshots of the time-varying permeability field.}
  \label{Permstv}
\end{figure}

\begin{figure}
  \begin{center}
      \includegraphics[width=0.65\textwidth]{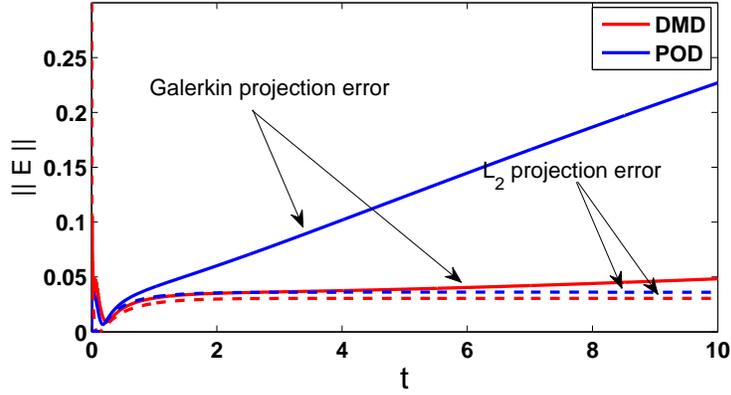}
  \end{center}
  \caption{Variations of the L$_2$ projection (represented by dashed lines) and Galerkin projection (represented by solid lines) errors with time: time-varying permeability field. Results are obtained using POD and DMD modes.}
  \label{Err_tv}
\end{figure}

Next, we consider a time-varying permeability field where the permeability
is changed in every time instant by a mobility function (c.f., two-phase
flow simulations \cite{rainer, eh09, eghe05}). At this stage, we use simplified mobility
functions to demonstrate that dynamic modes obtained from DMD can be used
to accurately predict flow dynamics for different initial conditions and
for time-varying permeability fields.
As such, we consider the following coefficient:
\begin{equation}\label{ktv}
\kappa_t(x,t)=\kappa(x)\;\Gamma(x,r(t)),
\end{equation}
where $\kappa$ is the coefficient of the permeability field shown in Figure \ref{Perm_incl1}, $\Gamma(x,r(t))$ is defined as
\begin{equation}
\Gamma(x,r(t))=\left\{
                 \begin{array}{ll}
                   2  \quad \mbox{if} \quad x \in \Pi\\
                   1  \quad \mbox{else}
                 \end{array}
               \right.
\end{equation}
and $\Pi$ is a circle of a time-varying radius $r(t)$ and center (0,0) and $r(t)=10\;t$. Figure \ref{Permstv} shows the permeability field
at three different instants. In this case, the matrix A is time-dependent and then it is evaluated at each time step in Equation (\ref{AE}). Similar to the previous analysis, we follow POD- and DMD-based approaches and derive reduced-order models to investigate their appropriateness for time-varying porous media problems. In Figure \ref{Err_tv}, we plot the variations of the L$_2$ projection and Galerkin projection errors with time. We observe small  L$_2$ projection errors when using POD and DMD modes. However, a large error is reached when projecting the governing equations onto the space spanned by POD modes to obtain a reduced-order model. This error keeps growing as time evolves. On the other hand, a small error is obtained when using DMD modes. This indicates the suitability of the use of DMD modes for model reduction of flows in time-varying and highly heterogeneous porous media.

\subsection{Parameter-dependent case}

\begin{figure}
  \begin{center}
      \subfigure[Case 1]{\includegraphics[width=0.32\textwidth]{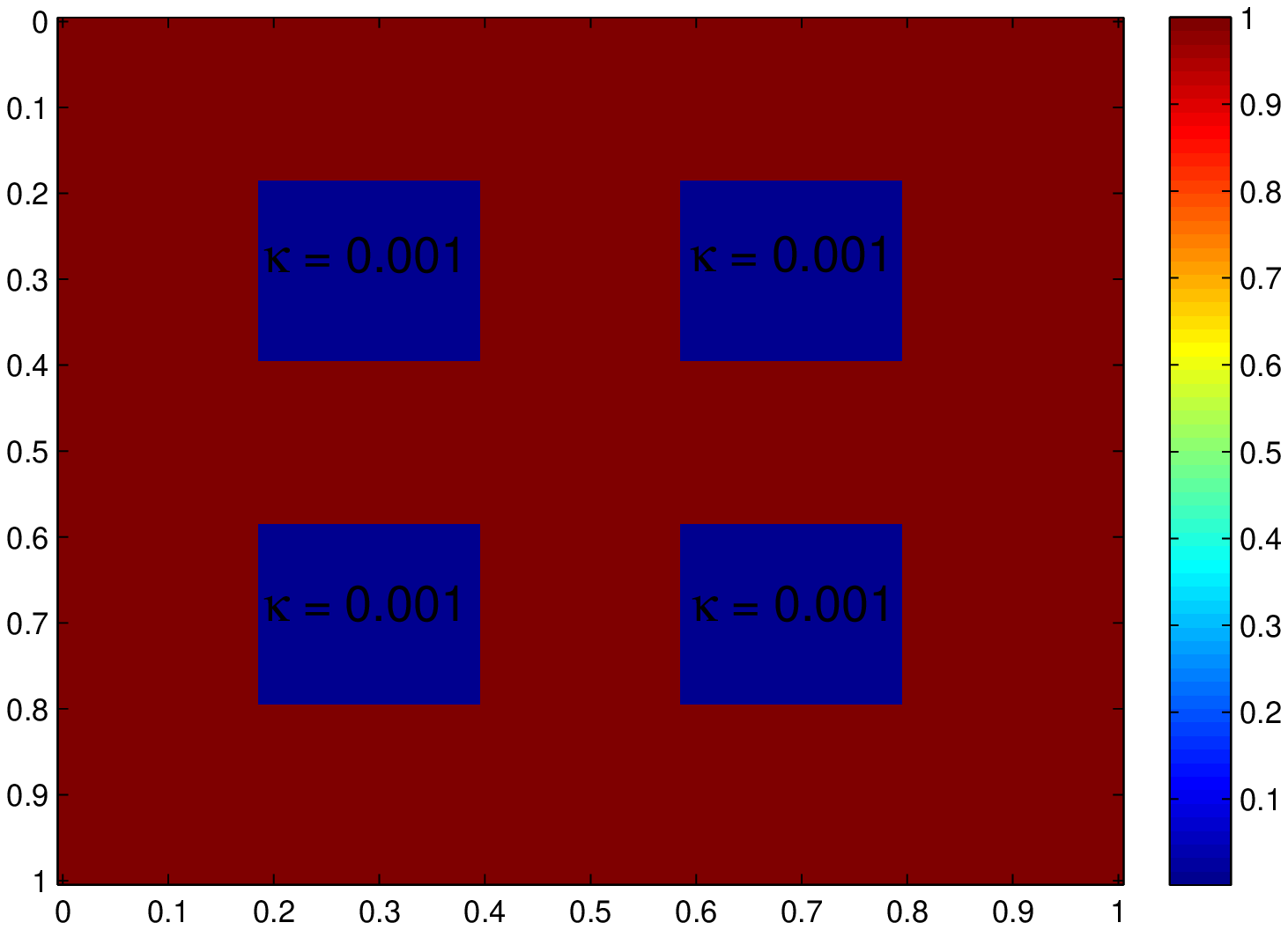}\label{1}}
      \subfigure[Case 2]{\includegraphics[width=0.3\textwidth]{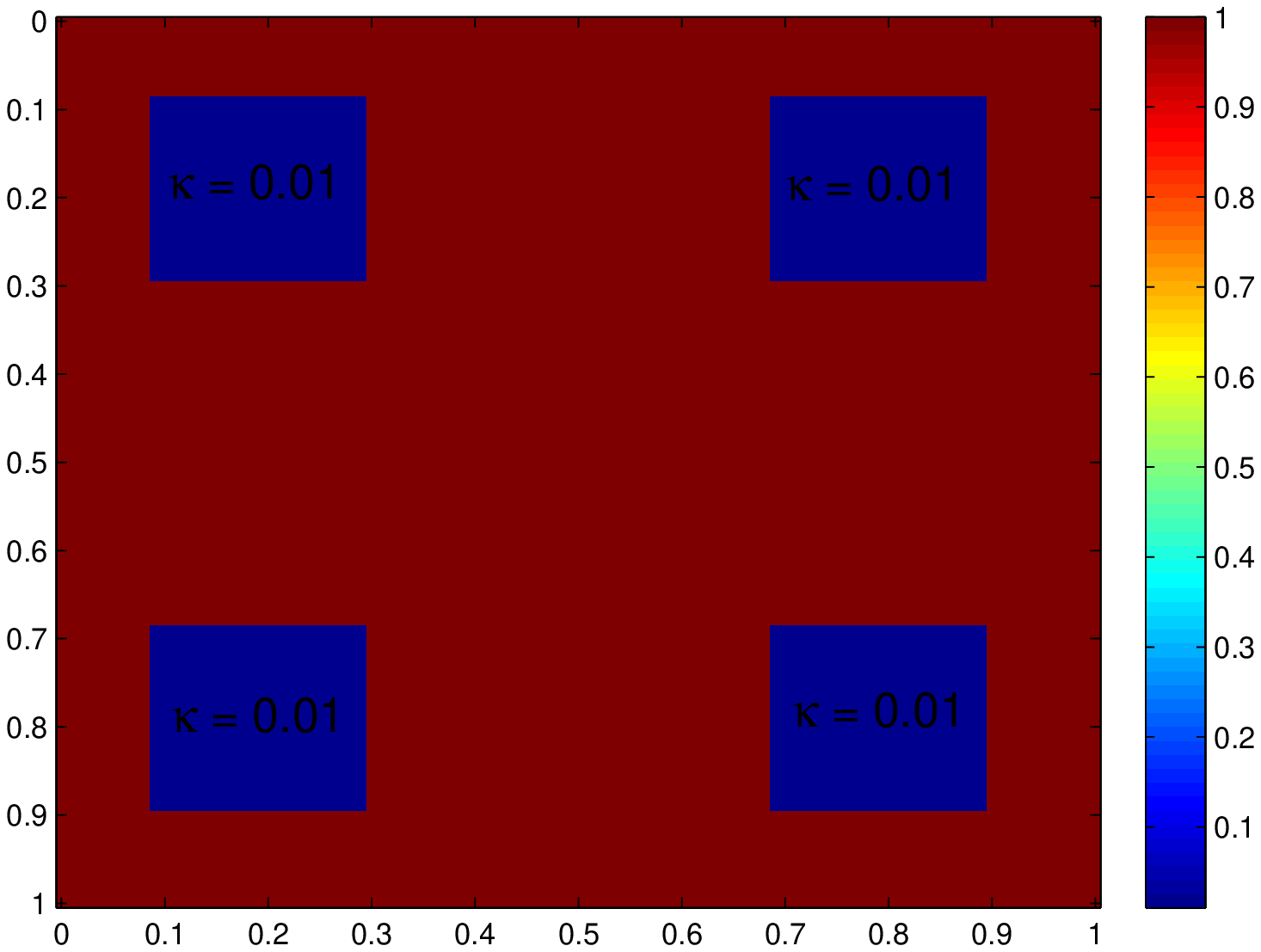}\label{2}}
      \subfigure[Case 3]{\includegraphics[width=0.3\textwidth]{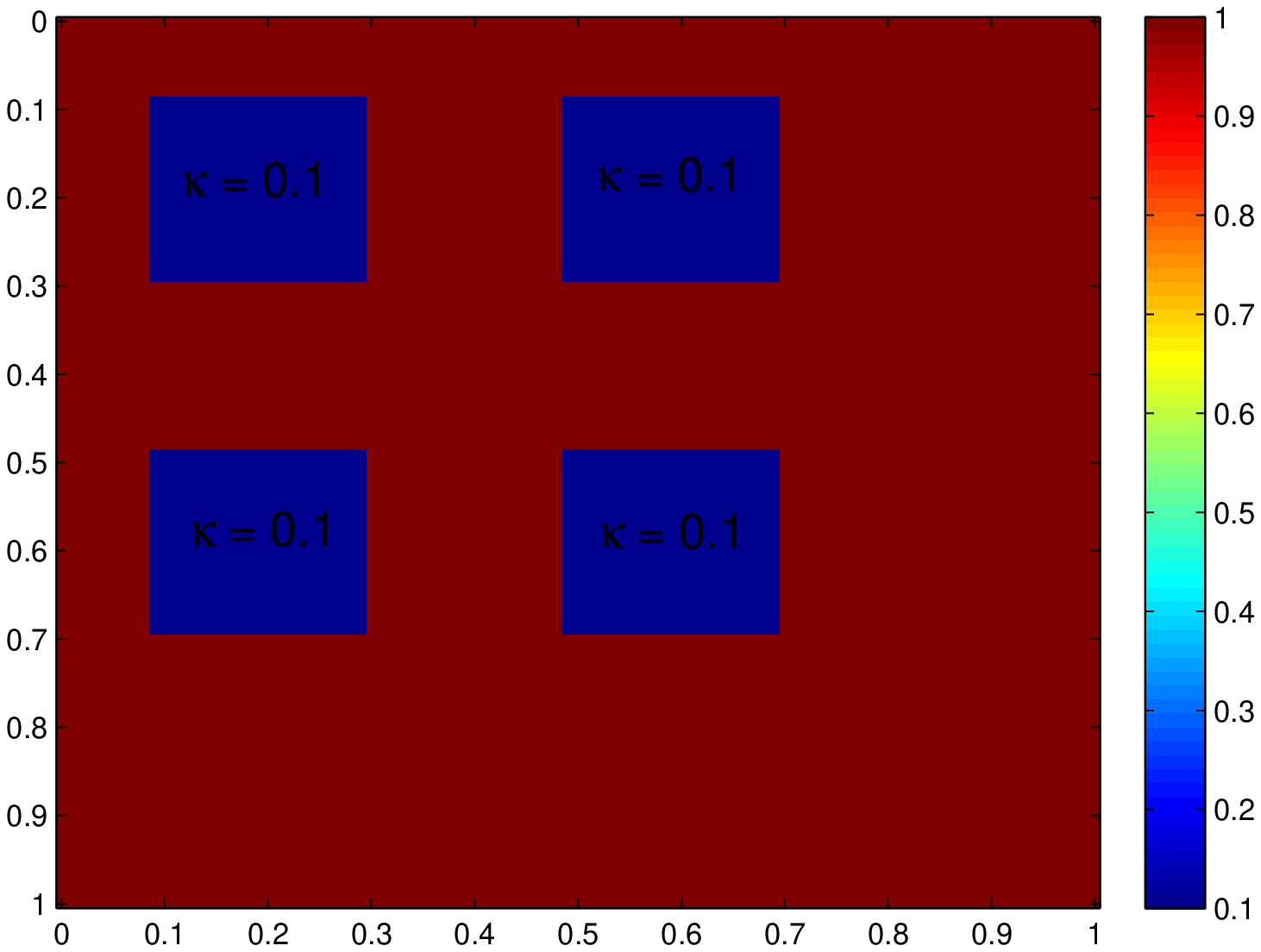}\label{3}}
      \subfigure[Case 4]{\includegraphics[width=0.3\textwidth]{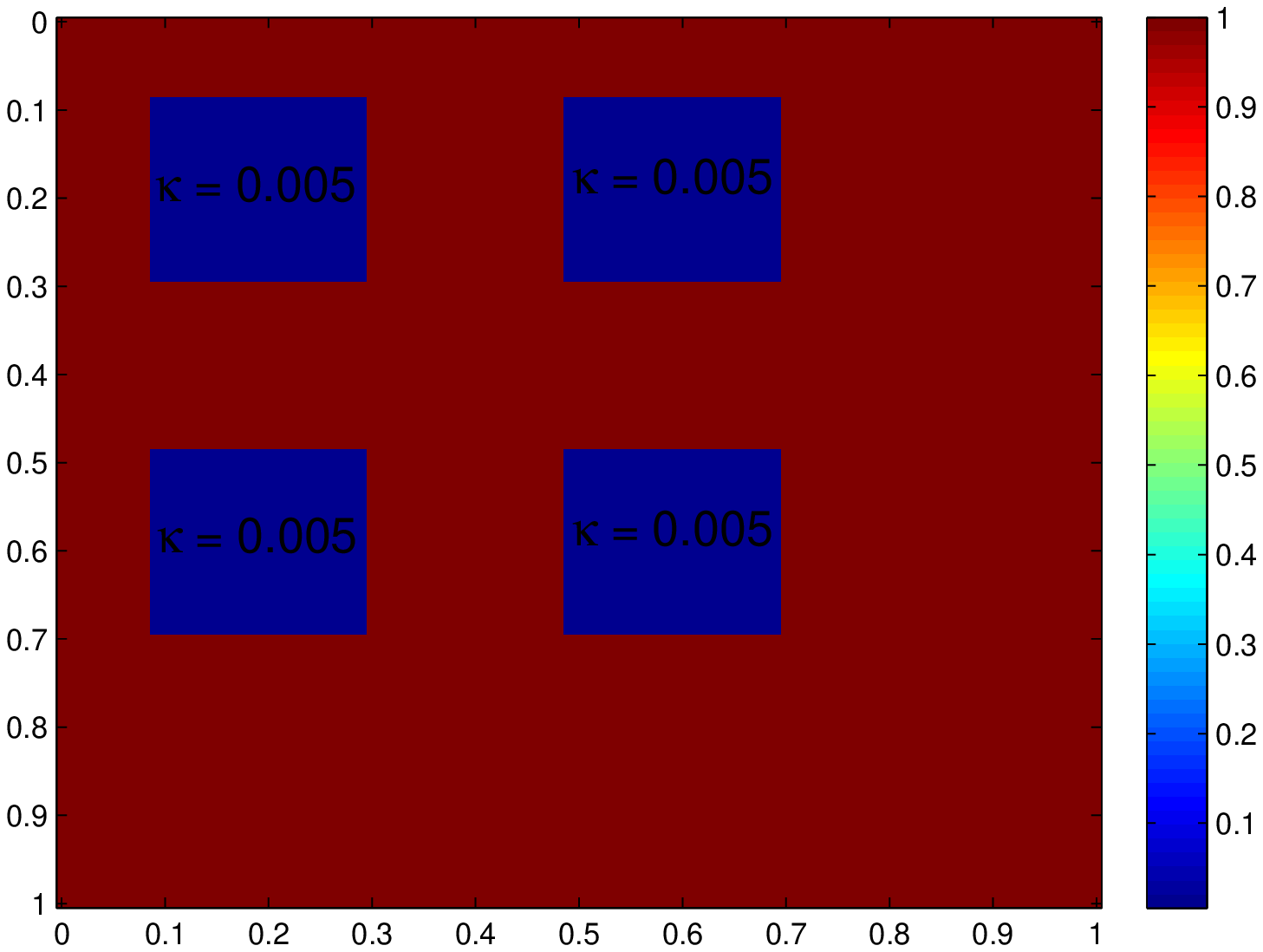}\label{4}}
      \subfigure[Case 5]{\includegraphics[width=0.3\textwidth]{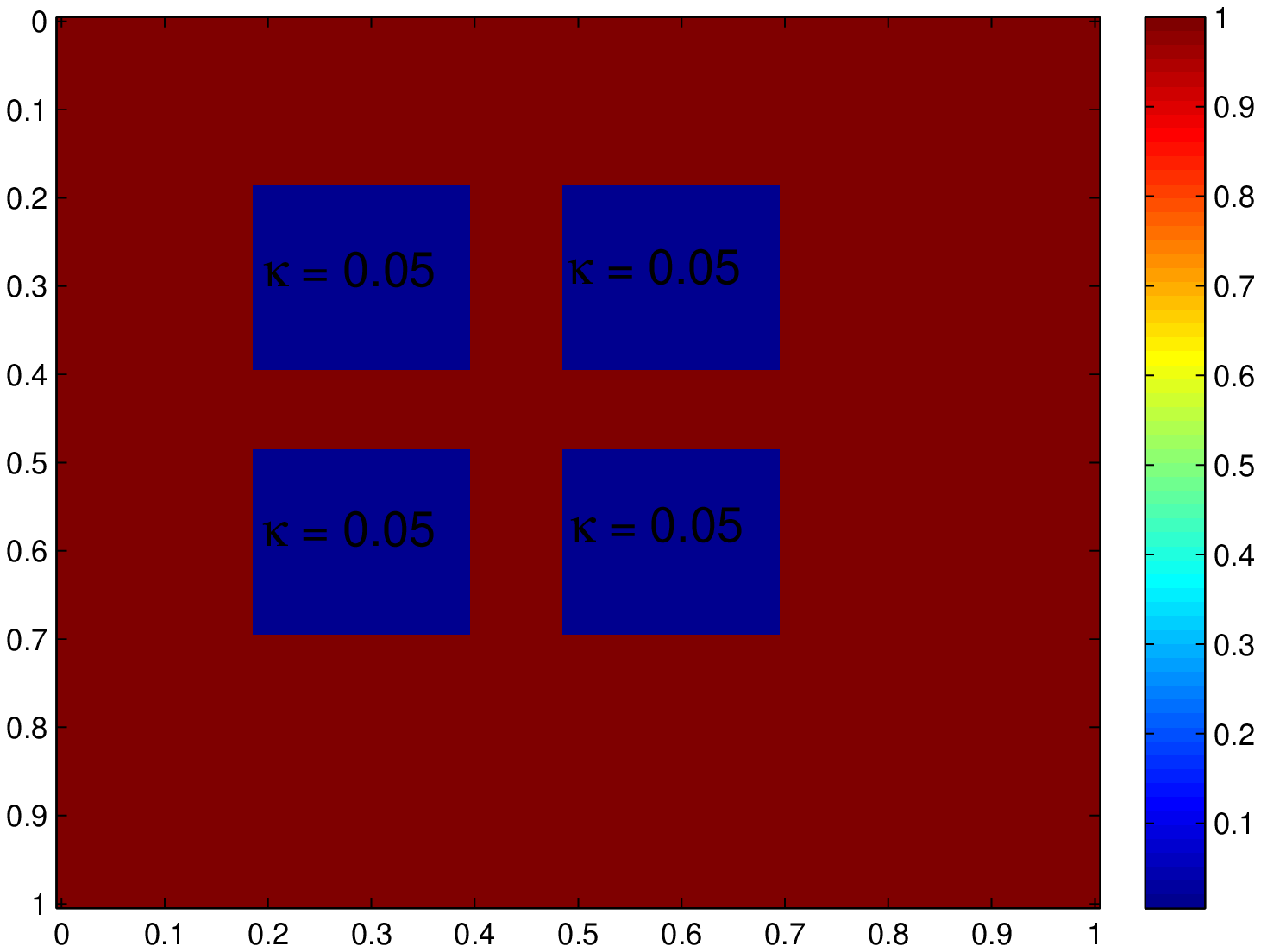}\label{5}}
  \end{center}
  \caption{Five permeability fields each containing inclusions with different contrast.}
  \label{Perms}
\end{figure}

\begin{figure}[ht]
 \begin{center}
 \includegraphics[width=0.65\textwidth]{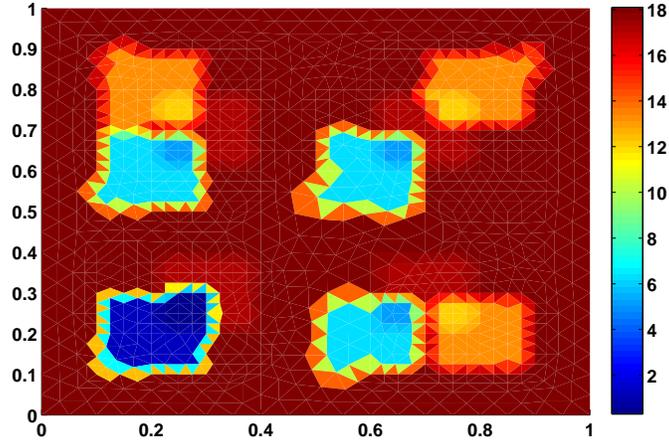}
 \end{center}
\caption{Permeability field (a linear combination of five different permeability fields shown in Figure \ref{Perms}).}\label{Perm_par_dep}
\end{figure}

\begin{figure}[ht]
 \begin{center}
 \includegraphics[width=0.65\textwidth]{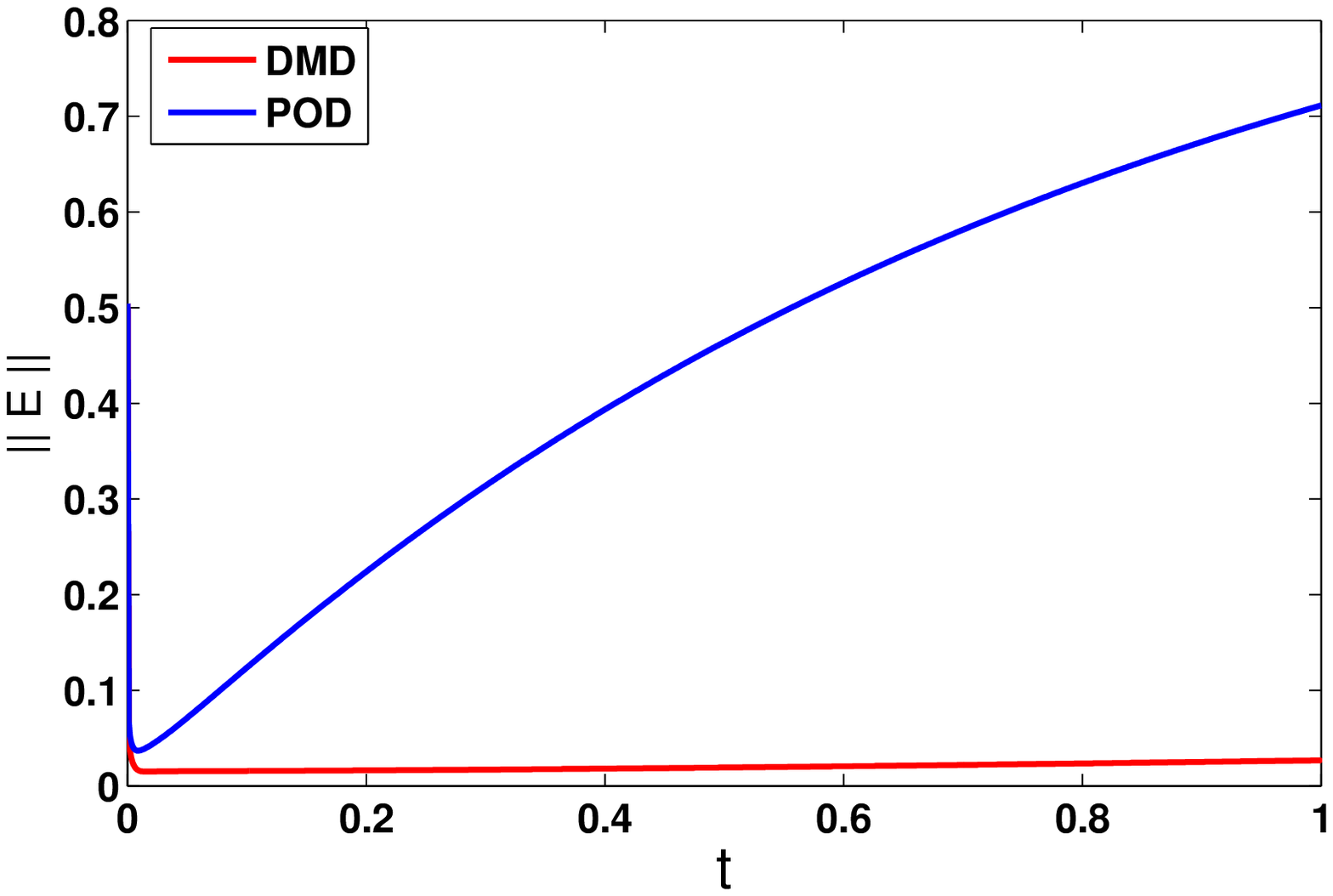}
 \end{center}
\caption{Variations of the Galerkin projection error with time. Results are obtained using POD and DMD modes.}\label{Error_par_dep}
\end{figure}

In this section, we investigate the robustness of model reduction techniques with respect to moderate variations in the permeability distribution, the contrast, the initial conditions, and the forcing inputs. We first consider a permeability field which is represented by a linear combination of five different permeability fields with each containing low-conductivity inclusions. These permeability fields are shown in Figure \ref{Perms} where each permeability field contains low-conductivity inclusions in different locations. This, for example, corresponds to the case where locations of low conductivity regions are not deterministic. The coefficient describing the resulting permeability is expressed as
\begin{equation}\label{perm}
\kappa(x;\mu)=\mu_1 \kappa_1(x) + \mu_2 \kappa_2(x) + \mu_3 \kappa_3(x) + \mu_4 \kappa_4(x)+ \mu_5 \kappa_5(x)
\end{equation}
The resulting permeability field obtained for $\{\mu_1,\mu_2,\mu_3,\mu_4,\mu_5\}=\{1,5,2,10,0.1\}$ is depicted in Figure \ref{Perm_par_dep}. We compute the POD and DMD modes for each of the five permeability configurations (shown in Figure \ref{Perms}) and collect them in a global matrix as
\begin{equation}\label{perm}
\Phi_{global}=\Big\{\Phi^1,\cdots,\Phi^5\Big\}.
\end{equation}
Then, we derive a general reduced-order model, with $5 \times m$ dimensions, by Galerkin projecting the governing Equation (\ref{Eq17}) onto the space spanned by the POD and DMD global modes and check its capability to predict accurately the case shown in Figure \ref{Perm_par_dep}. The temporal variations of the Galerkin projection error are plotted in Figure \ref{Error_par_dep}. Different initial conditions are considered and similar trends are observed. Unlike POD, DMD predicts the flow field with good accuracy. An error of 2\% is obtained. This error is comparable to the error between the reference and approximate solutions obtained when using the DMD modes computed directly for the permeability field shown in Figure \ref{Perm_par_dep}. These results show the robustness of the global DMD modes for developing reduced-order models that can be efficiently used to analyze the sensitivity of the dynamical behavior of the flow to moderate variations in the structure of permeability field (in terms of distribution and contrast).

\begin{figure}[ht]
 \begin{center}
 \includegraphics[width=0.65\textwidth]{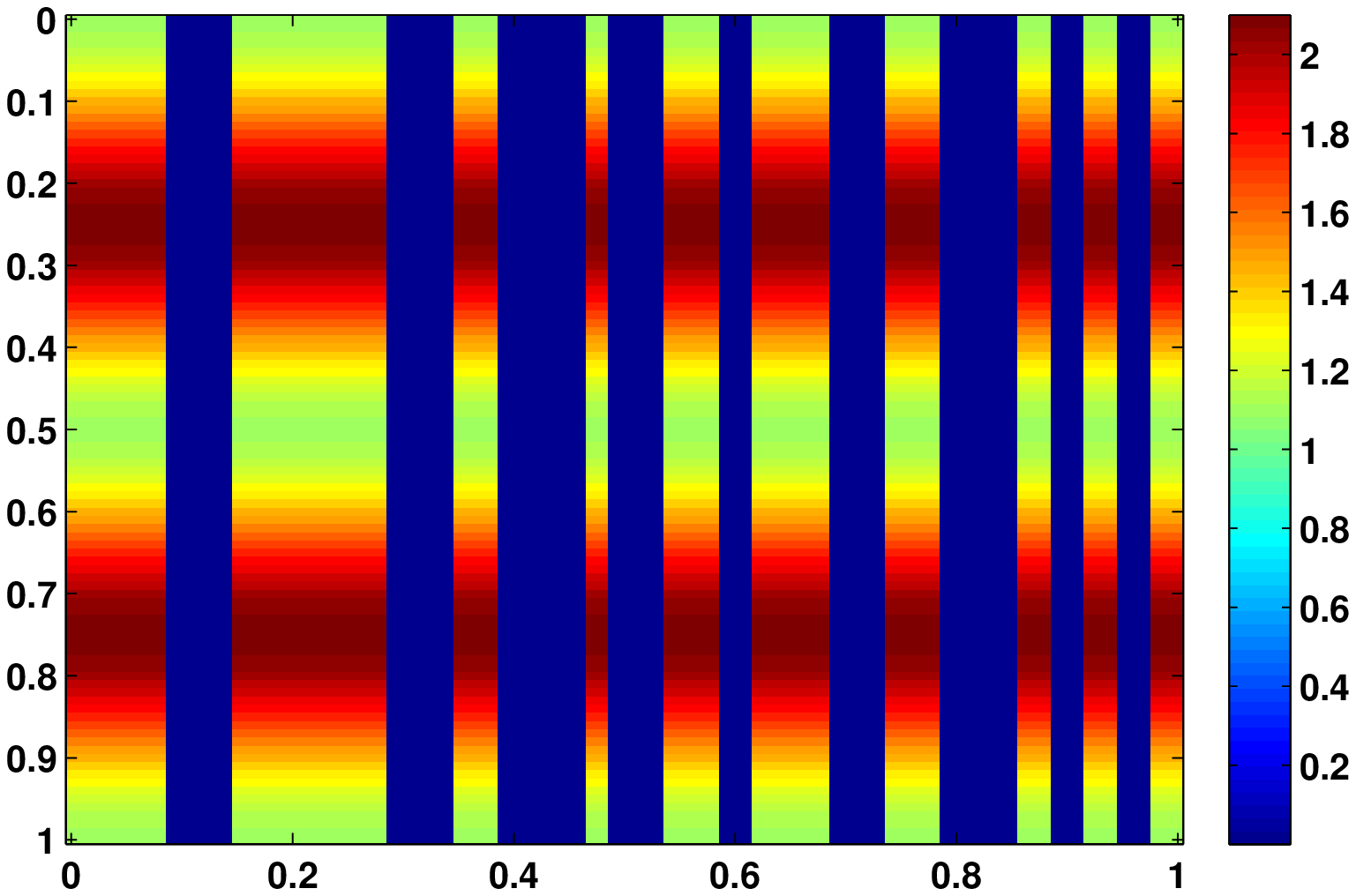}
 \end{center}
\caption{Permeability field obtained for $\epsilon =1$ and $f=100$.}\label{Perm_channel_smooth}
\end{figure}

\begin{figure}
  \begin{center}
      \subfigure[DMD]{\includegraphics[width=0.45\textwidth]{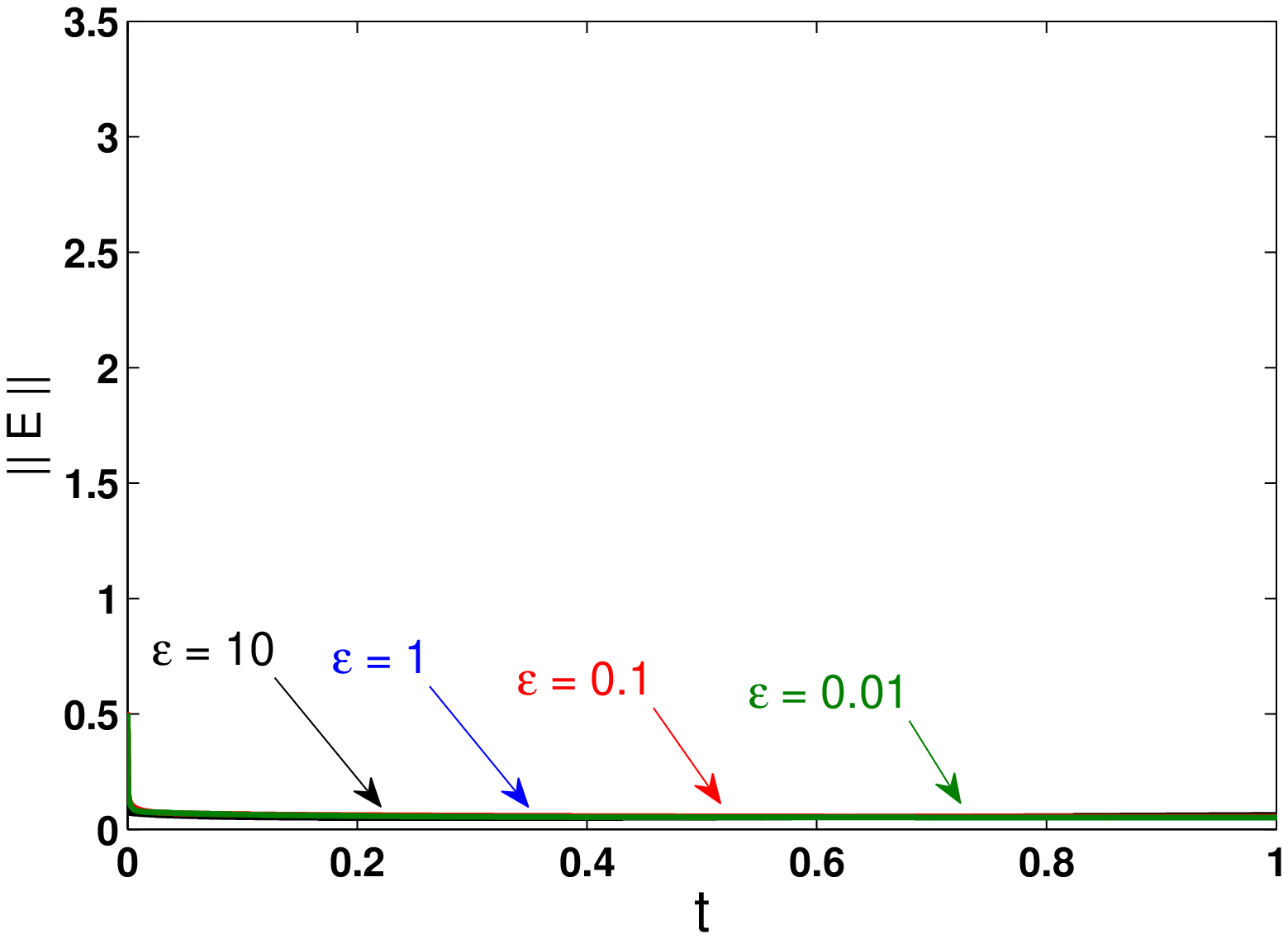}\label{ErrorDMD_smooth}}
      \subfigure[POD]{\includegraphics[width=0.45\textwidth]{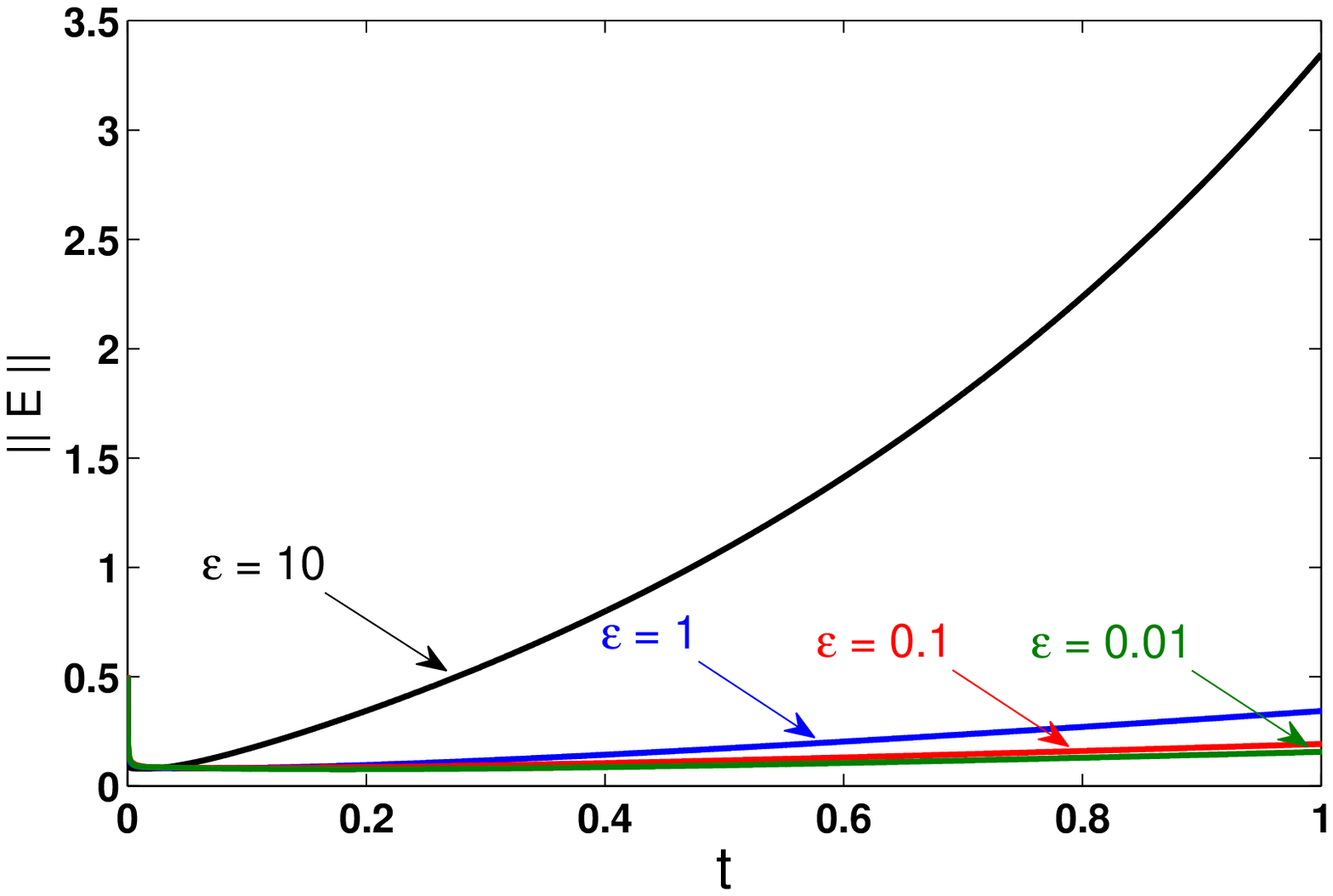}\label{ErrorPOD_smooth}}

  \end{center}
  \caption{Variations of the Galerkin projection error with time. Results are obtained using POD and DMD modes.}
  \label{Error_smooth}
\end{figure}

To investigate further the suitability of POD and DMD modes to model flows in varying and highly heterogeneous porous media, we consider the channel-type permeability (shown in Figure \ref{Perm_cha1}) and multiply its coefficient by a smooth positive spatial function; that is,
\begin{equation}\label{perm_smooth}
\kappa_s(x;\epsilon;f)=\kappa(x)\times(1+\epsilon+\sin(2\pi f x)\sin(2\pi f y) ).
\end{equation}
The obtained permeability field for $\epsilon =1$ and $f=100$ is depicted in Figure \ref{Perm_channel_smooth}. We use POD and DMD modes generated for the permeability field shown in Figure \ref{Perm_cha1} and employ the Galerkin projection to obtain a reduced-order model which is used to predict the flow field resulting from the modified permeability field described by Equation (\ref{perm_smooth}). In Figure \ref{Error_smooth}, we plot the temporal variations of the projection error obtained from the POD- and DMD- based representations while varying the value of $\epsilon$. Large Galerkin projection errors are obtained when using POD modes. These errors increase substantially as the value of $\epsilon$ increases. This indicates that POD-based reduced-order model can be only valid for the original configuration considered when computing the modes. On the other hand, the DMD-based model reduction approach seems to be much less sensitive to variations in the permeability. In fact, it shows a great capability to predict the flow field as can be deduced from the small error values shown in Figure \ref{ErrorDMD_smooth} (about 8\% for different values of $\epsilon$).

\begin{figure}[ht]
 \begin{center}
 \includegraphics[width=0.65\textwidth]{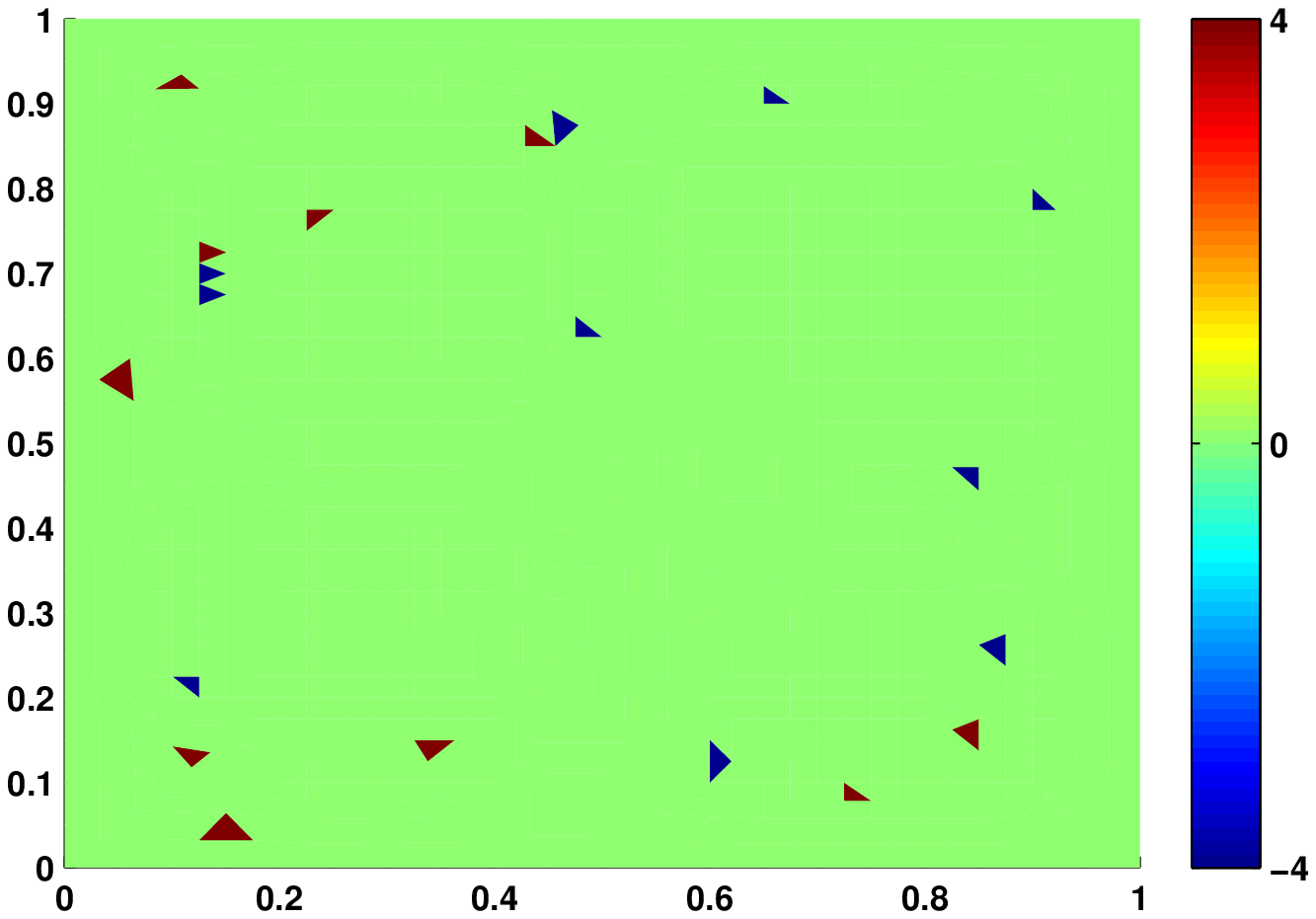}
 \end{center}
\caption{Spatial variations of the forcing $f$ over the domain $\Omega$.}\label{forcing2}
\end{figure}

\begin{figure}
  \begin{center}
      \includegraphics[width=0.65\textwidth]{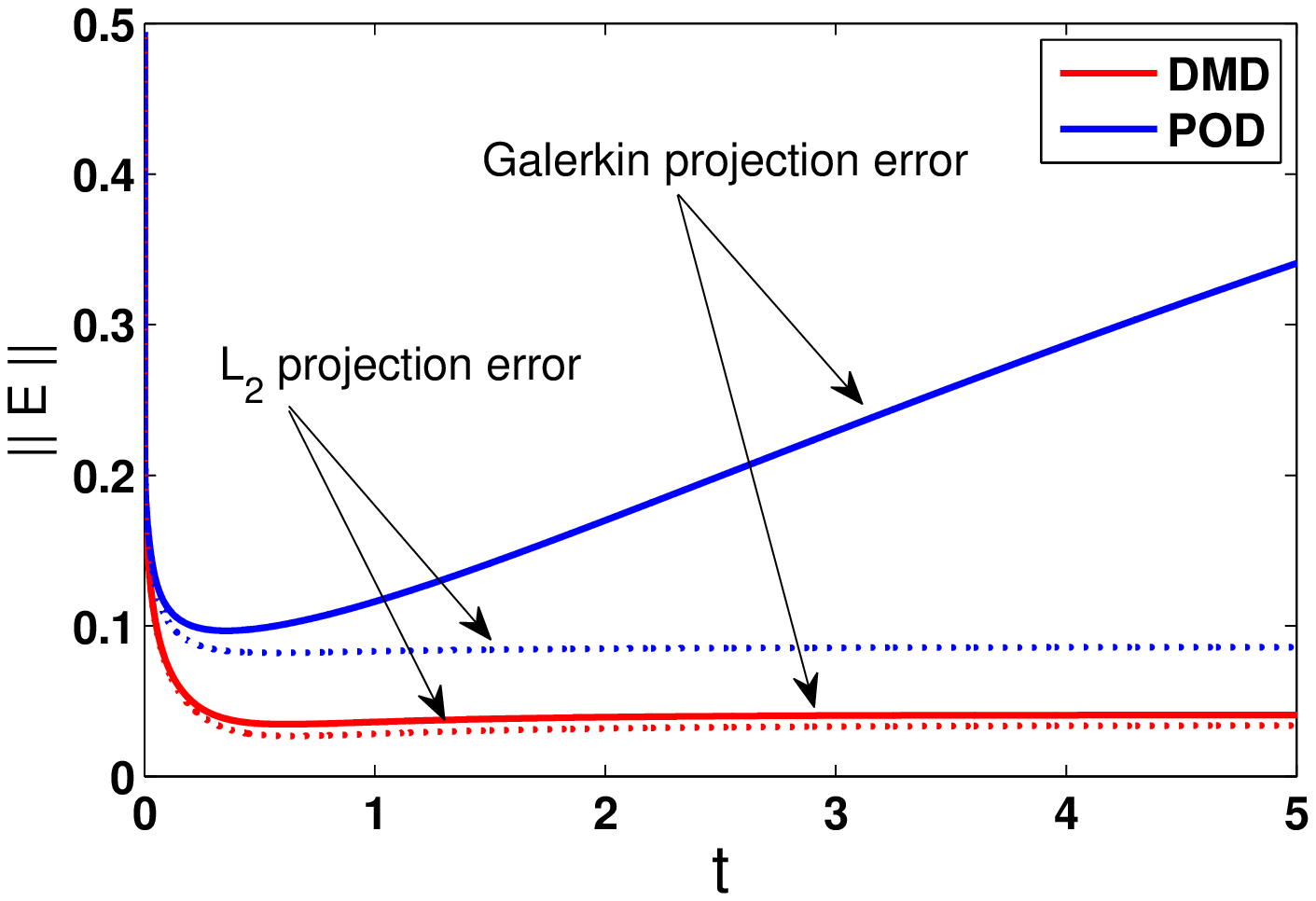}
  \end{center}
  \caption{Variations of the L$_2$ projection (represented by dashed lines) and Galerkin projection (represented by solid lines) errors with time. Results are obtained using POD and DMD modes.}
  \label{Err_Forcing}
\end{figure}

Next, we modify moderately the distribution and magnitude of the forcing term as shown in Figure \ref{forcing2}. We use the DMD and POD modes
obtained for the forcing case shown in Figure \ref{forcing} to derive a reduced-order model and predict the behavior of the flow field subjected to the modified forcing input. The variations of the L$_2$ projection and Galerkin projection errors with time are plotted in Figure \ref{Err_Forcing}. As expected, smaller errors are obtained from the L$_2$ projection in comparison to those obtained when using the reduced-order model. The use of DMD modes yields small errors. This shows the robustness of DMD-based approach to derive a reliable reduced-order model while moderately varying forcing inputs.

\section{Conclusions}

In this work, we applied proper orthogonal decomposition (POD) and dynamic mode decomposition (DMD) to flow in highly heterogeneous porous media with high contrast to derive a reduced-order model. Different numerical examples of flows in porous media characterized by highly varying permeability fields were considered. These permeability fields include channels and inclusions of high and low conductivity. The long-time dynamics of these flows are due to complex changes within low permeability regions. Through our cases, we investigated the capability of POD and DMD to capture the main flow characteristics and predict the flow field within a certain accuracy. The DMD-based approach showed better capability to reproduce the flow field when compared to the POD-based approach. This is mostly due to the DMD's ability to extract the dynamic information and particularly the modes that govern the long-time dynamics. We also considered parameter-dependent problems to investigate the robustness of the POD and DMD modes with respect to variations in the initial conditions, permeability field, and input forcing. We found that DMD-based approach provides robust basis functions to make accurate predictions of the dynamical behavior of flow in highly heterogeneous porous media.

\bibliography{RefDec2012}
\bibliographystyle{elsarticle-num}

\end{document}